# Contributions to the Generalized Coupon Collector and LRU Problems

**Christian BERTHET**
STMicroelectronics, Grenoble, France,

**Abstract**. Based upon inequalities on Subset Probabilities, proofs of several conjectures on the Generalized Coupon Collector Problem (i.e. CCP with unequal popularity) are presented. Then we derive a very simple asymptotic relation between the expectation of the waiting time for a partial collection in the CCP, and the Miss rate of a LRU cache.

**Address all correspondence to:** BERTHET Christian; E-mail: Christian.berthet@st.com

# 1. Introduction

It is known that Coupon Collector Problem (CCP) for a general popularity and a partial collection, on the one hand, and Miss rate computation of Least Recently Used (LRU) caches on the other hand are twin problems [Flajolet92]. The latter problem is also referred to as the Move-to-front search cost.

In the Coupon Collector Problem, a set of N, N>1, distinct objects (coupons, items...) is sampled with replacement by a collector in a way which is independent of all past events. This random process is frequently labelled Independent Reference Model (IRM). Each drawing produces item 'i' from the reference set of N items with probability $p_i$ such that $\sum_{i=1}^{N} p_i = 1$. Distribution $\{p_i\}$ is often called 'popularity'. Also, as in [Boneh97] we use the shorthand EL ('equally likely') to denote a uniform distribution (i.e. $p_i=1/N$).

CCP problem comes down to define how many trials ('waiting time') are needed before one has collected N items for the **complete** collection and a number n, n<N, for a **partial** collection of n items.

### (i) Organization of the document

The main aim of the current report is to show that, for a given popularity over N elements $\{p_i\}$, $1 \leq i \leq N$, and assuming that IRM hypothesis holds, the expectation (E) of the waiting time for a partial collection of size j in the CCP and the Miss rate (MR) of a LRU cache of size j on a stream of accesses belonging to a set of items with popularity $\{p_i\}$, are related in a very simple way.

When j increases, both quantities are asymptotically such that MR[j]*ΔE(j)≈1, where ΔE(j)=E(j+1)-E(j) is the forward finite difference operator applied to the expectation of the Waiting Time variable for a partial collection of size j.

This relation is proved in Section 6. Prior to this, the document is organized as follows.

Sections 2 and 3 are devoted to the definition of the two types of variables generally used in stating the CCP problem. We give some properties and formulas of these variables, as well as some open questions.

In Section 4 are introduced two recurrence relations on CCP probabilities which are new to our knowledge as well as some explicit values of these probabilities.

In Section 5 we prove that, compared to any non-uniform distribution, EL and only EL is maximal for the Cumulative Density Function (CDF) probability of the waiting time for a partial collection (and consequently EL and only EL is minimal for the Complementary CDF (CCDF) probability and expectation of the CCP Waiting Time).

In Section 6, we address the asymptotic relation between CCP and LRU. Section 7 is a conclusion and Section 8 details the references. Appendices with the proofs are in Section 9 to 15.

### (ii) Three main reference works

The main references we are using are listed in chronological order.

First comes [VonSchelling54] who gave an early version of the exact expression of probability and expectation of waiting time for a partial CCP and a non-uniform distribution.

Then [Flajolet92] gives a variant of Von Schelling formula for expectation together with another expression using an integral notation (which we will not deal with in this report).

[Boneh97] gives a thorough and detailed review of CCP and some conjectures that are addressed hereafter. They also give some approximations of the waiting time expectation which are not considered here.

### (iii) Notation

We assume a probability law with general distribution $\{p_i\}$, $i \in \{1..N\}$, $N > 1$, $1 > p_i > 0$, and $\sum_{i=1}^{N} p_i = 1$. For a subset J of the reference set $\{1,..,N\}$, we use uppercase $P_J$ to denote the probability of subset J, i.e. the sum of probabilities of the elements of J: $P_J = \sum_{i \in J} p_i$.

## 2. T_n Variable: Waiting Time

### (i) Probability Formula

We use the definition of [Boneh97] for the Waiting Time. This variable is "*Tj*: The number of drawings needed to complete a sub-collection of size *j*".

In the following we use the notation: 'n' is the sub-collection size, out of 'N' possible items in the reference set, and 'k' the number of drawings.

The formula for the probability of this variable (pdf form) was first given by Von Schelling [VonSchelling54] (using a somewhat different notation), $N \geq n \geq 1$, $k \geq 1$:

$$\boxed{\Pr[T_n = k] = \sum_{j=0}^{n-1} (-1)^{n-1-j} \binom{N-j-1}{N-n} \sum_{|J|=j} P_J^{k-1}(1 - P_J)}$$

Cumulative form (so-called CDF): $\Pr[T_n \leq k] = \sum_{0 < l \leq k} \Pr[T_n = l]$ and complementary cumulative form (CCDF): $\Pr[T_n > k] = 1 - \Pr[T_n \leq k]$ are easily computed.

It can be proved that $\Pr[T_n = k]$ is always null for k<n (see [Berthet17]) and that:

$\Pr[T_n = n] = n! \sum_{|J|=n} \prod_{j \in J} p_j$ and in particular, for n=N: $\Pr[T_N = N] = N! \prod_{i=1}^{N} p_i$.

### (ii) Uniform case

For EL (using [Boneh97] denomination), expression of this probability is very simple using Stirling numbers of 2nd order, which counts the number of ways to partition k objects into n non-empty blocks : $S(k,n) = \frac{1}{n!} \sum_{i=0}^{n} (-1)^{n-i} \binom{n}{i} i^k$ with S(n,n)=1, $\forall n \geq 0$, and S(k,n)=0 when k<n.

Probability is: $\Pr[T_n = k] = \frac{N! S(k-1, n-1)}{(N-n)! N^k}$ for an incomplete collection and

$\Pr[T_N = k] = \frac{N! S(k-1, N-1)}{N^k}$ for a complete collection.

Stirling numbers of the 2nd kind verify the recurrence relation: $S(k,n) = n S(k-1,n) + S(k-1,n-1)$. This leads to : $S(k,n) = \sum_{m=1}^{k} S(m-1, n-1) \cdot n^{k-m}$, which is obtained by enumerating recurrence relation S(m,n) with m varying from 1 to k, multiplying each side by $n^{k-m}$, adding the rows and considering that S(0,d)=0 for d>0.

Consequently a very simple expression exists for the CDF of a **complete** collection:

$\Pr[T_N \leq k] = \sum_{m \leq k} \frac{N! S(m-1, N-1)}{N^m} = \frac{N! S(k, N)}{N^k}$. Unfortunately no such simple CDF relation exists for an incomplete collection.

Also, for EL case, stemming directly from the recurrence relation on Stirling numbers there exists a recurrence relation on probabilities:

$$\Pr[T_{n+1}=k+1]=\frac{N-n}{N}\Pr[T_n=k]+\frac{n}{N}\Pr[T_{n+1}=k]$$, which means that, at step k+1, either one gets a new item with probability 1-n/N since there were n items at step k, otherwise (with probability n/N), there were already n+1 items in the collection.

No similar recurrence relation exists for a non-EL distribution.

### (iii) Waiting Time Expectation: Flajolet&al. and VonSchelling notations

It is direct that $\sum_{k\geq 0}\Pr[T_n>k]=\sum_{j=0}^{n-1}(-1)^{n-1-j}\binom{N-j-1}{N-n}\sum_{|J|=j}\sum_{k\geq 0}P_J^k$, hence $T_n$ expectation is:

$E[T_n]=\sum_{j=0}^{n-1}(-1)^{n-1-j}\binom{N-j-1}{N-n}\sum_{|J|=j}\frac{1}{1-P_J}$. This is the expression given by Flajolet&al (Formula 14a of the reference paper [Flajolet92]) for a partial collection.

For the full collection, formula is: $E[T_N]=\sum_{j=0}^{N-1}(-1)^{N-1-j}\sum_{|J|=j}\frac{1}{1-P_J}$.

This notation is equivalent to the "Von Schelling notation" using index change k=N-j:

$E[T_n]=\sum_{k=N-n+1}^{N}(-1)^{n-1-N+k}\binom{k-1}{N-n}\sum_{|J|=N-k}\frac{1}{1-P_J}=\sum_{k=N-n+1}^{N}(-1)^{n-1-N+k}\binom{k-1}{N-n}\sum_{|J|=k}\frac{1}{P_J}$ which appears in

[Schelling54], and: $E[T_N]=\sum_{k=1}^{N}(-1)^{k-1}\sum_{|J|=k}\frac{1}{P_J}$ for the complete collection.

It is easy to see that, for EL and complete collections, since $\sum_{|J|=k}\frac{1}{P_J}=\binom{N}{k}\frac{N}{k}$, then:

$E[T_N]=NH_N$, where $H_N$ is the Nth harmonic number, thanks to the remarkable equality: $\sum_{p=1}^{m}(-1)^{p-1}\binom{m}{p}\frac{1}{p}=\sum_{p=1}^{m}\frac{1}{p}=H_m$. Formula is generalized to partial collections of EL probabilities $E[T_n]=N(H_N-H_{N-n})$ (Appendix Section 9 gives a possible proof).

### (iv) Pdf curves

Following graph shows the pdf probability (decimal log) of a complete collection for N=12, k≥12 and three different popularities: uniform, Zipf and generalized Zipf (aka power-law) with 0.5 parameter.

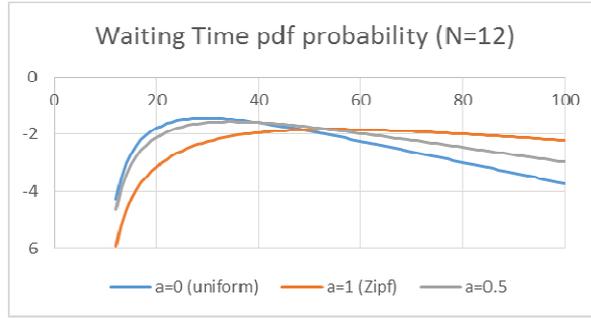

For k=N (initial point of each curve) pdf probability is respectively: $\frac{N!}{N^N}$, $H_N^{-N}$ and $H_{N,a}^{-N}(N!)^{1-a}$ for uniform, Zipf law and generalized Zipf law with skewness 'a' where $H_N$ (resp. $H_{N,a}$) is the Nth (resp. Generalized) Harmonic number.

We verified that up to N=100 (so it is conjectured for N>100) that the abciss of the pdf maximum for an EL distribution is o(N*lnN), i.e. the same trend as the expectation but slightly below, since $H_N$=ln(N)+γ+o(1/N), where γ is the Euler-Mascheroni constant (γ≈0.5772). For any other power-law distribution, abciss of the maximum is not known.

### (v) Observation in EL case

Let us consider the CDF expression $\Pr[T_N \leq k]$ extended to the case where k (k>N) is not integer, i.e. belongs to the continuous domain. Then we define $\Pr[T_N \leq E[T_N]]$, i.e. the probability that waiting time for a complete collection is less or equal to its expectation.

Calculating this expression using definition of Stirling numbers of the 2$^{nd}$ order (again assuming extension to the continuous domain, with the first parameter of the Stirling number not integer any more) produces for EL the following figure (blue curve) : $\Pr[T_N \leq NH_N]$, plotted for N up to 1000. It is compared to $Ne^{-H_N}$ (grey curve), whose limit, when N→∞, is $e^{-\gamma} \approx 0.56146$. Limit of $\Pr[T_N \leq E[T_N]]$ is slightly above $e^{-\gamma}$ value.

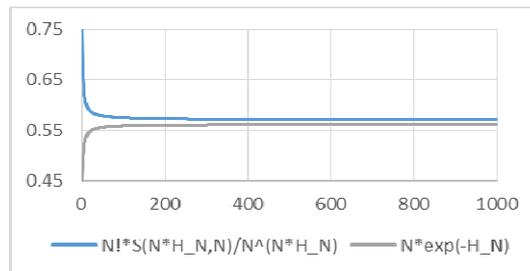

It stands that: $\sum_{i=0}^{N}(-1)^{N-i}\binom{N}{i}\left(\frac{i}{N}\right)^{NH_N}$ can be approximated by $\sum_{i=0}^{N}(-1)^i\binom{N}{i}e^{-iH_N}$.

On the one hand, $1-\frac{i}{N} \approx e^{-\frac{i}{N}}$ when i <<N, and on the other hand, terms $\left(1-\frac{i}{N}\right)^{NH_N}$ and $e^{-iH_N}$ both tend to 0 when i gets closer to N.

It follows that: $\lim_{N \to \infty} \Pr[T_N \leq NH_N] = \lim_{N \to \infty} (1 - e^{-H_N})^N = e^{-e^{-\gamma}} \approx 0.570376$.

This limit was given in Erdos and Renyi 1961 seminal paper [Erdos61], formula (27) in the form: $\lim_{N \to \infty} \Pr[T_N < N(\ln N + \gamma)] = e^{-e^{-\gamma}}$.

### (vi) An Empirical Observation for power-law Popularities

Following graph shows both CDF curves and expectation values of the waiting time variable for a complete collection (N=15), and different values of the parameter 'a' of a power-law, a=0, 1, 2, 3, 4. Computations are carried out using Von Schelling formulas.

It is striking that, regardless of the value of 'a', *CDFs calculated at the value of the expectation* are systematically ~ 0.6.

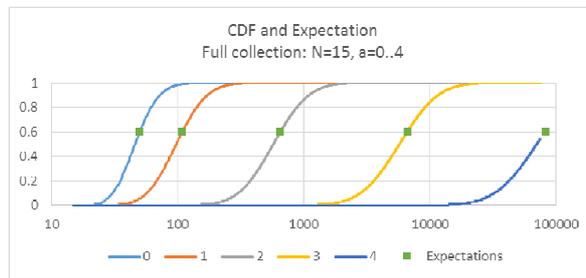

In other words, for a given N, expression $\Pr[T_N \leq E[T_N]]$ looks like a constant or so, whatever the power-law parameter value. We did check it as well for a =0.1,..,0.9.

The ~.6 value obtained for EL case (i.e. a=0) and N=15 is coherent with the graph of previous paragraph.

We conjecture that as long as N is large enough, this observation is true for any power-law popularity (and possibly, any non-EL popularity ?). Thus CDF expression $\Pr[T_N \leq k]$ for $k = E[T_N]$ is very similar to the value obtained for EL popularity, and consequently, when N increases: $\Pr[T_N \leq E[T_N]] \approx e^{-e^{-\gamma}} \approx 0.5703$, regardless of the power-law parameter value.

# 3. $W_k$ Variable: Working Set

A second variable is defined in [Boneh97]: "*Yn* :The number of different items observed in the first *n* drawings". In the following we change the Y letter to a W, which stands for Working Set and this will be explained later.

The variable $W_k$ denotes a probability function as a function of the number of distinct items and not a function of the number of trials as the previous variable.

We use the same notation (n among N after k trials) as before.

### (i) Probability

After [Flajolet92, formula 6, p214], Boneh relates the probability of the two variables [Boneh97, formula (27)]: $\Pr[W_k < n] = \Pr[T_n > k]$, or: $\Pr[W_k \geq n] = \Pr[T_n \leq k]$.

Then, using formula of $T_n$ probability, $W_k$ pdf probability is:

$$\Pr[W_k = n] = \Pr[T_{n+1} > k] - \Pr[T_n > k] = \sum_{j=0}^{n}(-1)^{n-j}\binom{N-j-1}{N-(n+1)}\sum_{|J|=j}P_J^k - \sum_{j=0}^{n-1}(-1)^{n-1-j}\binom{N-j-1}{N-n}\sum_{|J|=j}P_J^k,$$

i.e: $$\boxed{\Pr[W_k = n] = \sum_{j=0}^{n}(-1)^{n-j}\binom{N-j}{N-n}\sum_{|J|=j}P_J^k}.$$

Obviously, $\Pr[W_k = n] = 0$ for n>N and k<n.

Again, using Von Schelling representation of $T_n$ and relation $\Pr[W_k < n+1] = \Pr[T_{n+1} > k]$,

$W_k$ CDF probability is: $\Pr[W_k \leq n] = \sum_{j=0}^{n}(-1)^{n-j}\binom{N-j-1}{N-n-1}\sum_{|J|=j}P_J^k$.

Noticing that for N≥n>0, $\sum_{j=0}^{n}(-1)^{n-j}\binom{N-j-1}{N-n-1}\sum_{|J|=j}1 = \sum_{j=0}^{n}(-1)^{n-j}\binom{N-j-1}{N-n-1}\binom{N}{j} = 1$ (see Appendix 1 in [Berthet17]), the CCDF form is: $\Pr[W_k > n] = \sum_{j=0}^{n}(-1)^{n-j}\binom{N-j-1}{N-n-1}\sum_{|J|=j}\left(1 - P_J^k\right)$.

Let us stress that for the complete collection: $\Pr[W_k \leq N] = \sum_{j=0}^{N}(-1)^{N-j}\binom{N-j-1}{-1}\sum_{|J|=j}P_J^k$, binomial coefficient is 1 for j=N, and 0 elsewhere, hence $\Pr[W_k \leq N] = 1$ or $\Pr[W_k > N] = 0$. Also: $\Pr[W_k = N] = \sum_{j=0}^{N}(-1)^{N-j}\binom{N-j}{0}\sum_{|J|=j}P_J^k = (-1)^N\sum_{j=0}^{N}(-1)^j\sum_{|J|=j}P_J^k$. As shown in [Berthet17] Corollary 1, this expression is null for k<N.

### (ii) EL case

For EL, the subset-of-subsets rule on binomial coefficients leads to a very simple expression: $\Pr[W_k = n] = \binom{N}{n}\frac{1}{N^k}\sum_{j=0}^{n}(-1)^{n-j}\binom{n}{j}j^k = \frac{N!S(k,n)}{(N-n)!N^k}$ with Stirling numbers of 2nd order. It follows, for a complete collection: $\Pr[W_k = N] = \frac{N!S(k,N)}{N^k}$.

Note that, $W_k$ pdf formula is the same as $T_n$ CDF formula for a complete collection. As for $W_k$ CDF probability, there is no simple expression.

### (iii) Expectation

Expectation of $W_k$ is: $E[W_k] = \sum_{n \geq 0} P(W_k > n) = \sum_{j=0}^{N-1} \left( \sum_{n=j}^{N-1} (-1)^{n-j} \binom{N-j-1}{N-n-1} \right) \sum_{|J|=j} \left(1 - P_J^k\right)$ obtained by commuting the summations.

Noting that: $\sum_{n=j}^{N-1} (-1)^{n-j} \binom{N-j-1}{N-n-1} = \sum_{u=0}^{N-j-1} (-1)^u \binom{N-j-1}{u} = 1_{j=N-1}$ where $1_A = 1$ when A is true else 0, finally: $E[W_k] = \sum_{|J|=N-1} (1 - P_J^k) = \sum_{i=1}^{N} \left(1 - (1-p_i)^k\right)$. This is formula (26) of [Boneh97].

Remarkably $E[W_k]$ is also the average Working Set function [Fagin77] used in caching analysis and this is the reason why we use for $W_k$ variable the same patronyme.

### (iv) Summary of Formulas

Notation: (n different items among N possible after k trials)

| Variable | $T_n$ | $W_k$ |
|---|---|---|
| Definition | Number of drawings needed to complete a sub-collection of size n in k trials | Number of different items observed in the first *k* drawings |
| Pdf | $\Pr[T_n = k] = \sum_{j=0}^{n-1} (-1)^{n-1-j} \binom{N-j-1}{N-n} \sum_{|J|=j} P_J^{k-1}(1-P_J)$ | $\Pr[W_k = n] = \sum_{j=0}^{n} (-1)^{n-j} \binom{N-j}{N-n} \sum_{|J|=j} P_J^k$ |
| CDF | $\Pr[T_n \leq k] = \sum_{j=0}^{n-1} (-1)^{n-1-j} \binom{N-j-1}{N-n} \sum_{|J|=j} (1 - P_J^k)$ | $\Pr[W_k \leq n] = \sum_{j=0}^{n} (-1)^{n-j} \binom{N-j-1}{N-n-1} \sum_{|J|=j} P_J^k$ |
| CCDF | $\Pr[T_n > k] = \sum_{j=0}^{n-1} (-1)^{n-1-j} \binom{N-j-1}{N-n} \sum_{|J|=j} P_J^k$ | $\Pr[W_k > n] = \sum_{j=0}^{n} (-1)^{n-j} \binom{N-j-1}{N-n-1} \sum_{|J|=j} \left(1 - P_J^k\right)$ |
| Expectation | Also called Waiting time for a n-size partial collection | Also called WS(k) function (average 'working set') |
| | $E[T_n] = \sum_{j=0}^{n-1} (-1)^{n-1-j} \binom{N-j-1}{N-n} \sum_{|J|=j} \frac{1}{1-P_J}$ | $E[W_k] = \sum_{j=1}^{N} (1 - (1-p_j)^k)$ |

Note that all probability expressions as well as $T_n$ expectation can be transformed with a summation index change (N-j).

### (v) $W_k$ Recurrence Relation for EL

For an EL distribution, [Read98] gives a recurrence relation for the variable $W_k$. With his notation, probability is $p_n(r+1, s+1) = [(n-s) p_n(r,s) + (s+1) p_n(r, s+1)] / n$

r = 1, 2, 3,... ; s = 1, ... , min (r, n).

which is justified: "*to find (s + 1) different types of card in (r + 1) packets, either we previously had s different types in the first r packets and then (with chance (n - s)/n)*" *found one of the other (n - s) types in the (r + 1)th packet, or we already had (s + 1) different types in the first r packets and then got one of these types again in the (r + l)th packet (which happens with chance (s + 1)/n)*".

Rewritten with our notation (with k substituted for r, N for n and n for s) $\Pr[W_{k+1} = n+1] = \frac{N-n}{N}\Pr[W_k = n] + \frac{n+1}{N}\Pr[W_k = n+1]$. This recurrence stems directly from recurrence on Stirling numbers. Let's note that [Read98] does not distinguish between the two variables giving $W_k$ probability in formula (5) and $T_N$ expectation in formula (6).

Also let us mention that, again, there is no such recurrence for non-EL popularities.

### (vi) Properties

Relation between $T_n$ CDF and $W_k$ pdf for **complete** collection of EL popularity extends itself to any popularity, i.e.: $\Pr[W_k = N] = \sum_{j=0}^{N}(-1)^{N-j}\sum_{|J|=j}P_J^k = 1 - \Pr[T_N > k] = \Pr[T_N \leq k]$.

A consequence of formula (27) [Boneh97] is: $\Pr[T_n = k] = \Pr[W_{k-1} < n] - \Pr[W_k < n]$

which implies also: $\Pr[T_{n+1} = k] - \Pr[T_n = k] = \Pr[W_{k-1} = n] - \Pr[W_k = n]$.

Similarly: $\Pr[W_k = n] = \Pr[T_{n+1} > k] - \Pr[T_n > k]$, n<N.

For k=n, i.e., the lucky case each trial produces a new type of coupon, both probabilities are equal: $\Pr[W_n = n] = \Pr[T_{n+1} > n] - \Pr[T_n > n] = 1 - \Pr[T_n > n] = \Pr[T_n \leq n] = \Pr[T_n = n]$.

### (vii) Other Relations

From previous relations, it follows that: $\sum_{q=1}^{n-1}\Pr[W_k = q] = \Pr[T_n > k]$, and by complementarity: $\sum_{q=n}^{N}\Pr[W_k = q] = \Pr[T_n \leq k]$. We have also: $\sum_{u>k}\Pr[T_n = u] = \Pr[W_k < n]$.

Another relation is: $\sum_{n=1}^{N}\Pr[T_n = k] = \sum_{n=1}^{N}\Pr[W_k > n-1] - \sum_{n=1}^{N}\Pr[W_{k-1} > n-1] = E[W_k] - E[W_{k-1}]$

hence:

$$\boxed{\sum_{n=1}^{N}\Pr[T_n = k] = \sum_{i=1}^{N}p_i(1-p_i)^{k-1}}.$$

This sum is always positive for any k>0, and for EL case: $\sum_{n=1}^{N}\Pr[T_n = k] = \left(1 - \frac{1}{N}\right)^{k-1}$.

A dual relation is: $\sum_{k\geq 0}\Pr[W_k < n] = \sum_{k\geq 0}\Pr[T_n > k] = E[T_n]$, which is equivalent to:

$$\boxed{\sum_{k\geq 0}\Pr[W_k = n] = E[T_{n+1}] - E[T_n], \ n < N}.$$

For an EL distribution, this is: $\sum_{k \geq 0} \Pr[W_k = n] = N(H_{N-n} - H_{N-n-1}) = \dfrac{N}{N-n}$.

### (viii) Proof of Correctness of CCP Probability

From $T_n$ pdf formula, we have: $\Pr[T_1 = k] = 1_{k=1}$, and $\Pr[T_2 = k] = \sum_{i=1}^{N} p_i^{k-1}(1-p_i) \cdot 1_{k>1}$.

Proving that Probability formula is correct in the sense that it is always positive: $\Pr[T_n = k] > 0, \forall n : 1 \leq n \leq N, \forall k : k \geq n$ is not so obvious. For example, the proof of $\Pr[T_3 = k] > 0, \forall k \geq 3$ is equivalent to proving: $\sum_{|J|=2} P_J^{k-1}(1-P_J) > (N-2)\sum_{i=1}^{N} p_i^{k-1}(1-p_i)$, for any popularity of size N.
In the next Section, we prove this relation in the complete case, after introducing a specific quantity that makes expressions somehow less painful to manipulate.

## 4. Calculation on Sums of Subsets Probabilities

### (i) Definition

For a given popularity of size N, we introduce the notation: $\boxed{R_N^k = \sum_{0 \leq |J| \leq N} (-1)^{|J|} P_J^{\ k}}$.

Let us stress that higher index of $R_N^k$ is a convention and not an exponent.

$T_N$ probability for a **complete** collection is easily derived:

$$\Pr[T_N > k] = \sum_{|J|=0}^{N-1}(-1)^{N-1-|J|} P_J^{\ k} = 1 - (-1)^N R_N^k, \text{ and then: } \Pr[T_N \leq k] = (-1)^N R_N^k.$$

### (ii) Properties

It stands that: $R_N^0 = R_N^1 = \ldots R_N^{N-1} = 0$ and $R_N^N = (-1)^N N! \prod_{1 \leq j \leq N} p_j$.

The following property holds: $\sum_{0 \leq |J| \leq N}(-1)^{|J|}(a + P_J)^N = R_N^N, \quad \forall a \text{ real},$ which stems directly from the binomial development and nullity of $R_N^k$ when k<N.

### (iii) A "full history" Recurrence Relation

From binomial decomposition: $\sum_{|J|=j}(1 - P_J)^k = \sum_{u=0}^{k}(-1)^u \binom{k}{u}\sum_{|J|=j} P_J^{\ u}$, and using index change (j=N-j), it holds that: $R_N^k = \sum_{0 \leq j \leq N}(-1)^{N-j} \sum_{u=0}^{k}(-1)^u \binom{k}{u} \sum_{|J|=j} P_J^{\ u}$, i.e.:

$R_N^k = \sum_{u=N}^{k}(-1)^u \binom{k}{u} \sum_{0 \leq j \leq N}(-1)^{N-j} \sum_{|J|=j} P_J^{\ u}$. Hence: $\boxed{R_N^k = \sum_{u=N}^{k}(-1)^{N+u} \binom{k}{u} R_N^u}$.

This expression leads to a tautology when N+k is even (in particular for k=N).

For k=N+1, one has: $R_N^{N+1} = (N+1)R_N^N - R_N^{N+1}$, so: $\boxed{R_N^{N+1} = \frac{N+1}{2} R_N^N}$.

More generally, **if k is odd**: $\boxed{R_N^{N+k} = \frac{1}{2} \sum_{u=N}^{N+k-1}(-1)^{N+u} \binom{N+k}{u} R_N^u}$.

Otherwise (i.e., k is even): $\sum_{u=N}^{N+k-1}(-1)^{N+u} \binom{N+k}{u} R_N^u = 0$. This leads to a dual expression

**when k is odd**: $R_N^{N+k} = \sum_{u=N}^{N+k-1}(-1)^{N+u} \frac{1}{u}\binom{N+k}{u-1} R_N^u = \frac{1}{(N+k+1)} \sum_{u=N}^{N+k-1}(-1)^{N+u} \binom{N+k+1}{u} R_N^u$.

### (iv) Another Recurrence on CCP probability

From definition: $R_N^k = \sum_{0 \leq |J| \leq N} (-1)^{|J|} P_J^k$, we introduce the following notation on a new distribution $\left\{ q_j = \dfrac{p_j}{1 - p_l} \right\}, j \in (1..N), j \neq l$, defined on the reference set of size (N-1) resulting from the exclusion of element 'l': $R_{N-1,\{l\}}^u = \sum_{\substack{0 \leq |J| \leq N-1 \\ l \notin J}} (-1)^{|J|} \left( \dfrac{P_J}{1 - p_l} \right)^u$.

It holds that: $R_{N-1,\{l\}}^{N-1} = (-1)^{N-1} (N-1)! \prod_{1 \leq j \leq N, j \neq l} \dfrac{p_j}{1 - p_l} = \dfrac{(-1)}{N \cdot p_l (1 - p_l)^{N-1}} R_N^N$.

In a previous paper [Berthet17], we have shown the following relation on subsets probabilities (Relation 3, page 11): $\sum_{|J|=j} P_J^k (1 - P_J) = \sum_{l=1}^{N} \left( p_l \sum_{|J|=j, l \notin J} P_J^k \right), \forall k \geq 0$.

A direct consequence is: $\boxed{R_N^k = R_N^{k-1} - \sum_{l=1}^{N} p_l (1 - p_l)^{k-1} R_{N-1,\{l\}}^{k-1}, \forall k > 0}$.

In other words, $R_N^k$ expression can be iteratively obtained if the expression is known for a decremented size of the reference set and a correspondingly decremented exponent. Unfortunately, this does not give a clue for the 'initial condition' of the recurrence.

We can now define a recurrence relation on CCP Waiting Time probability:

Let $T_{N-1,\{l\}}$ be the variable related to the distribution $\left\{ q_j = \dfrac{p_j}{1 - p_l} \right\}, j \in (1..N), j \neq l$

defined on the reference set of size (N-1) resulting from the exclusion of element 'l'.

Then CCP probability verifies: $\boxed{\Pr[T_N = k] = \sum_{l=1}^{N} p_l (1 - p_l)^{k-1} \Pr[T_{N-1,\{l\}} \leq k - 1], \forall k > 0}$.

This implies that, for a complete collection, $\Pr[T_N = k]$ is always positive for k≥N, since, whatever the element 'l', $\Pr[T_{N-1,\{l\}} \leq k - 1] > 0$. This statement can be easily proven by recurrence on the size of the distribution set N.

### (v) Explicit expressions of $R_N^k$, k≥N

Here we derive explicit expressions of $R_N^k$ for k=N+1, k=N+2 and k=N+3, which are valid for any popularity.

As seen before, from binomial development, we have: $R_N^{N+1} = R_N^N \binom{N+1}{2} \dfrac{1}{N}$.

In Appendix Section 11 using the above recurrence, we prove by induction on the size N of the distribution that: $R_N^{N+2} = R_N^N \binom{N+2}{3} \frac{1}{4N}\left(3 + \sum_{i=1}^{N} p_i^2\right)$.

Using again binomial development, this leads to: $R_N^{N+3} = R_N^N \binom{N+3}{4} \frac{1}{2N}\left(1 + \sum_{i=1}^{N} p_i^2\right)$

So far, alas, we have not found a closed-form expression of $R_N^{N+4}$. A generalization to any exponent is an open question.

### (vi) Case of Uniform Popularity

For a uniform popularity, and replacing $R_N^k$ notation by $EL_N^k$, it stands from the definition of Stirling numbers that: $EL_N^k = \sum_{0 \le j \le N} (-1)^j \binom{N}{j} \left(\frac{j}{N}\right)^k = \frac{(-1)^N N! S(k,N)}{N^k}$.

A derivation can be obtained using the recurrence relation on Stirling numbers of 2nd order $S(n+k,n) - S(n+k-1,n-1) = n S(n+k-1,n)$. Then: $S(n+k,n) = \sum_{m=1}^{n} m S(m+k-1,m)$

One can obtain successively ( $EL_N^k$ are then derived):

$$S(N+1,N) = \sum_{m=1}^{N} m S(m,m) = \binom{N+1}{2}$$

$$S(N+2,N) = \sum_{m=1}^{N} m S(m+1,m) = \sum_{m=1}^{N} m \binom{m+1}{2} = \binom{N+2}{3} \frac{3N+1}{4}$$

$$S(N+3,N) = \sum_{m=1}^{N} m \binom{m+2}{3} \frac{3m+1}{4} = \binom{N+3}{4}\binom{N+1}{2}$$

$$S(N+4,N) = \binom{N+4}{5} \frac{15N^3 + 30N^2 + 5N - 2}{48}$$

$$S(N+5,N) = \binom{N+5}{6}\binom{N+1}{2} \frac{3N^2 + 7N - 2}{8}.$$

It appears that $S(n+k,n)\binom{n+k}{k+1}^{-1}$ is always a polynomial of degree k-1, obviously equal to 1 when n=1 and positive when n>1. This was stated by Griffiths [Griffiths12].

# 5. Proof of Conjectures

### (i) Boneh and Hofri 1997 Conjecture on the appearance of graphs

In the 1997 version (pg. 10) of their work, Boneh and Hofri [Boneh97] give the following statement: "We conjecture that the appearance of Fig. 1, where the **duration** curve lies entirely above the **detection** curve (except for the first two points: $t = 0, 1$, where they coincide) is unique to the EL case, and that in all other cases they intersect".

'Duration' is the inverse function of the Waiting Time Expectation and 'Detection' is the Working Set Expectation divided by N. A proof of this conjecture on the so-called duration and detection curves can be stated as follows.

Each of these curves is a sequence of segments. As stated by the authors, they coincide at (0,0) and (1,1/N). For x=2, 'Detection' function is $y = 1 - \frac{1}{N}\sum_{i=1}^{N}(1-p_i)^2 = \frac{1}{N}\left(2 - \sum_{i=1}^{N}p_i^2\right)$, hence its slope on [1,2] segment is: $\frac{1}{N}\left(1 - \sum_{i=1}^{N}p_i^2\right)$. Inverse of Waiting Time expectation has coordinates (x=E{$C_2$},y=2/N) with $E\{C_2\} = \left(\sum_{i=1}^{N}\frac{1}{1-p_i}\right) - (N-1)$, hence its slope on

$[1, E\{C_2\}]$ segment is: $\frac{1/N}{\left(\sum_{i=1}^{N}\frac{1}{1-p_i}\right) - N} = \frac{1}{N\left(\sum_{i=1}^{N}\frac{p_i}{1-p_i}\right)}$.

It is worth noting that in the uniform case, both slopes are equal to $(N-1)/N^2$.

For a non-uniform case, conjecture holds if it can be proved that the slope of "detection" function at point x=1$^+$ is steeper than that of the "Duration" function.

If this is the case, curves necessarily intersect later because, at the other end of the curve, "Detection" function is always strictly below 1, whereas inverse of waiting time function ends at point with coordinates (x= E[$T_N$],y=1).

This comes down to showing that $\left(1 - \sum_{i=1}^{N}p_i^2\right)^{-1} = 1 + \sum_{i=1}^{N}p_i^2 + \left(\sum_{i=1}^{N}p_i^2\right)^2 + ...$ is smaller than $\sum_{i=1}^{N}\frac{p_i}{1-p_i} = 1 + \sum_{i=1}^{N}p_i^2 + \sum_{i=1}^{N}p_i^3 + ..$ which is true if : $\sum_{i=1}^{N}p_i^k > \left(\sum_{i=1}^{N}p_i^2\right)^{k-1}, \forall k > 2$. This relation is proven in Appendix Section 10 Lemma 10.2 and completes the proof of the conjecture.

### (ii) EL duration and EL detection

Proving that in the EL (unifom case) duration lies above the detection is equivalent to proving that: $1 - \left(1 - \frac{1}{N}\right)^{N(H_N - H_{N-n})} \leq \frac{n}{N}, \forall n : 0 \leq n \leq N$, which is obvious for n=0, 1, and

N. For n=2: $1 - \frac{2}{N} \leq (1 - \frac{1}{N})^{\frac{2N-1}{N-1}}$ can be checked easily using WolframAlpha©. General case is proven by induction. Assuming relation is true for n, it holds for (n+1):

$(1 - \frac{1}{N})^{N(H_N - H_{N-(n+1)})} = (1 - \frac{1}{N})^{N(H_N - H_{N-n}) + \frac{N}{N-n}} \geq \left(1 - \frac{n}{N}\right)(1 - \frac{1}{N})^{\frac{N}{N-n}}$. Then it must be proved:

$\left(1 - \frac{n}{N}\right)(1 - \frac{1}{N})^{\frac{N}{N-n}} \geq \left(1 - \frac{n+1}{N}\right)$, i.e.: $(1 - \frac{1}{N})^N \geq \left(1 - \frac{1}{N-n}\right)^{N-n}, \forall n : 0 \leq n < N$, which is true because $(1-1/x)^x$ is increasing for x>=1.

### (iii) EL and only EL is maximal for Working Set Expectation

Let us consider the distribution $\left\{q_i = \frac{1 - p_i}{N - 1}\right\}$, 1≤i≤N. Obviously, it verifies: $\sum_{i=1}^{N} q_i = 1$ and is non-EL. Then, from $\sum_{i=1}^{N} q_i^k > \frac{1}{N^{k-1}}$ (see lemma 10.2 in Appendix Section 10), it holds that: $\sum_{i=1}^{N} \left(\frac{1 - p_i}{N - 1}\right)^k > \frac{1}{N^{k-1}}$, except for k=0,1 where equality holds.

Thus: $\sum_{i=1}^{N} (1 - p_i)^k > \frac{(N-1)^k}{N^{k-1}}$, or $\sum_{i=1}^{N} \left(1 - (1 - p_i)^k\right) < \sum_{i=1}^{N} \left(1 - \left(1 - \frac{1}{N}\right)^k\right)$. In other words, compared to all other non-uniform distributions, EL is always maximal regarding Working Set expectation for any k>1.

Relation $\sum_{i=1}^{N} (1 - p_i)^k > \sum_{i=1}^{N} \left(1 - \frac{1}{N}\right)^k$ has a number of consequences, for example, for k=3: $3\sum_{i=1}^{N} p_i^2 - \sum_{i=1}^{N} p_i^3 > \frac{3}{N} - \frac{1}{N^2}$.

### (iv) EL and only EL is minimal for Waiting Time Expectation

Boneh and Hofri have shown the minimality of EL compared to any other distribution regarding the expectation of the waiting time of a full collection. This result is extended in [Anceaume14] (Theorem 5 page 8) to any partial collection where they show that expectation of any non-EL popularity is higher or equal to that of EL (which does not mean **strictly** higher).

In Appendix Section 12, we use a different argument to show the **strict** minimality of EL regarding the expectation of Waiting Time $T_n$ for a **partial** collection.

### (v) EL and only EL is minimal for Waiting Time CCDF

In Appendix Section 13 we give a proof of minimality of EL w.r.t Waiting Time CCDF, first for a complete collection. We then give the sketch of the proof for a **partial** collection.

# 6. Relation between CCP Waiting Time Expectation and LRU Miss Rate

### (i) Uniform Distribution

In case of a uniform distribution (EL), expectation difference of the Waiting Time for a partial collection of size j is: $\Delta E[T_j] = \dfrac{N}{N-j}$ for 0≤j<N.

On the other hand, the miss rate of a LRU cache of size j for a uniform popularity on a support of size N verifies: $MR[j] = 1 - \dfrac{j}{N}$.

Therefore, an amazing relation holds between Waiting Time expectation difference and LRU miss rate: $MR[j] \cdot \Delta E[T_j] = 1$.

### (ii) General Distribution

In the general case of a non-EL probability, no such direct equality stands between the difference operator of expectation and the miss rate of an LRU cache for example expressed with King or Flajolet exact formulas (see [Berthet16]).

First let us remark that for j=1: $\Delta E[T_1] = \sum_{i=1}^{N} \dfrac{p_i}{1-p_i} = \sum_{i=1}^{N} \left( p_i (1 + p_i + p_i^2 + ..) \right)$ and

$MR[1] = 1 - \sum_{i=1}^{N} p_i^2$. Hence, relation $MR[1] \cdot \Delta E[T_1] = 1$ holds if and only if, for any k>2,

$\sum_{i=1}^{N} p_i^{k-1} \left( p_i - \sum_{j=1}^{N} p_j^2 \right) = 0$, and we know from lemma 10.3 (Appendix Section 10) that expression is null only for an EL distribution, otherwise it is strictly positive.

Therefore an extension to an arbitrary popularity of the relation between Expectation difference and LRU MR would make sense only as an asymptotic approximation when j and N increase.

There is a quite old result that defines an asymptotic approximation of LRU MR which is known to give excellent results for real-life values of cache size and support size.

### (iii) Fagin approximation of LRU miss rate

In 1977, an approximation of LRU miss rate was given by Fagin [Fagin77] under the classical IRM hypothesis. Fagin claims that "in a certain asymptotic sense" LRU miss rate can be approximated when the support size N increases, by an expression which (after moving to the continuous domain and assuming that popularity contains no large $p_i$, i.e., for all elements of the popularity: ln(1-$p_i$)=- $p_i$) we represent as:

$$\boxed{MR[j] \cdot \dfrac{d}{dj} WS^{-1}(j) \approx 1},$$

where WS$^{-1}$ is the inverse function of the Working Set function and '≈' denotes the "asymptotically close" condition. This approximation fell into oblivion during quite a few years before being recently rediscovered under the "Che approximation" label.

Of course it is not an exact calculation of LRU miss rate such as in King or Flajolet formulae, however, it is surprisingly precise and much valuable since it is computable on very large supports which is not the case for exact formulas.

In the next section, we show that there is another asymptotic relation, this time between the two variables of the CCP problem: $WS^{-1}(j) \approx E[T_j]$. Therefore, Fagin approximation can be extended by the following relation between the Miss Rate of an LRU cache of size j and the Waiting Time expectation for a partial collection of size j, when they are subject to the same popularity and obviously, the same IRM hypothesis:

$$\boxed{MR[j] \cdot \Delta E[T_j] \approx 1}.$$

### (iv) Algebraic proof of $WS^{-1}(j) \approx E[T_j]$

Intuitively, inverse function WS$^{-1}$(D) is the average of the size of a window containing D distinct addresses (or items,..) hence it is asymptotic to the expectation of CCP waiting time to collect a partial collection of D coupons.

It was noted, in [Boneh97] equation (57), and using our notation, that, for an EL distribution: $WS(E[T_N]) \approx N(1-e^{-H_N})$, i.e. "the expected number of items detected by the time the collector would expect to finish is extremely close to N, at N-e$^{-\gamma}$, with the shortfall essentially independent of N". Note that e$^{-\gamma}$ is the limit of $Ne^{-H_N}$ when N→∞.

In this section, first, we generalize Boneh observation and prove that this relation on T$_N$ (complete collection) is indeed true **for any popularity**, i.e:

$$\boxed{N > WS(E[T_N]) > N - e^{-\gamma}, \text{ for any popularity}}.$$

Then, we prove that a similar relation exists for a partial collection of size j, as long as j is large enough:    $\boxed{j > WS(E[T_j]) > j - e^{-\gamma}, \text{ for any popularity}}.$

Proofs are given in Appendix Section 14.

This leads us to the conclusion that asymptotically: $WS^{-1}(j) \approx E[T_j]$.

### (v) Application: Derivation of Doumas & Papanicolaou formulas for power-laws popularities.

The asymptotic approximation $E[T_j] \approx WS^{-1}(j)$ lends itself to a result obtained by A. Doumas and V. Papanicolaou [Doumas12] when the popularity distribution is a power-law (a.k.a. generalized Zipf law). This is described in Appendix Section 15.

# 7. Conclusion

In this document, we have shown a novel and very simple asymptotic relation between the Expectation difference of the CPP Waiting Time on the one hand, and the Miss rate of an LRU cache on the other hand, assuming that both are faced with the same popularity of items and same IRM hypothesis.

Prior to this, we have introduced inequalities on subsets probabilities, some of them were not already known. they allow for algebric proofs for some conjectures on the optimality of EL w.r.t to other popularities

# 9. Appendix: Expectation of partial collection for uniform distributions

We want to prove that for uniform distributions and a collection of size c among n items:

$$E[T_c] = \sum_{i=0}^{c-1}(-1)^{c-1-i}\binom{n-i-1}{n-c}\binom{n}{i}\frac{n}{n-i} = n(H_n - H_{n-c}).$$

From $\binom{n-i-1}{n-c}\binom{n}{i} = \binom{n}{c}\binom{c-1}{i}\frac{c}{n-i}$, one has: $E[T_c] = nc\binom{n}{c}\sum_{i=0}^{c-1}\binom{c-1}{i}\frac{(-1)^{c-1-i}}{(n-i)^2}$ or with a

summation index change: $E[T_c] = nc\binom{n}{c}\sum_{i=0}^{c-1}\binom{c-1}{i}\frac{(-1)^i}{(n-c+1+i)^2}$.

Known relation: $\int_0^1 x^a(1-x)^b\,dx = \frac{a!\,b!}{(a+b+1)!}$ obtained by iterated integration by parts is generalized to a variable bound, for a≥0, b≥0: $\int_0^u x^a(1-x)^b\,dx = \sum_{k=1}^{b+1}\frac{a!\,b!\,u^{a+k}(1-u)^{b+1-k}}{(a+k)!(b+1-k)!}$.

Then: $\int_0^1 \frac{1}{u}\left(\int_0^u x^a(1-x)^b\,dx\right)du = \int_0^1\sum_{k=1}^{b+1}\frac{a!\,b!\,u^{a+k-1}(1-u)^{b+1-k}}{(a+k)!(b+1-k)!}\,du = \frac{a!\,b!}{(a+b+1)!}\sum_{k=1}^{b+1}\frac{1}{(a+k)}$.

On the other hand, integral is also obtained by binomial expansion:

$$\int_0^u x^a(1-x)^b\,dx = \int_0^u\sum_{i=0}^{b}\binom{b}{i}(-1)^i x^{a+i}\,dx = \sum_{i=0}^{b}\binom{b}{i}\frac{(-1)^i}{a+i+1}u^{a+i+1}.$$

Hence: $\int_0^1 \frac{1}{u}(\int_0^u x^a(1-x)^b\,dx)du = \int_0^1\sum_{i=0}^{b}\binom{b}{i}\frac{(-1)^i}{a+i+1}u^{a+i}\,du = \sum_{i=0}^{b}\binom{b}{i}\frac{(-1)^i}{(a+i+1)^2}$

Thus we obtain the remarkable identity: $\sum_{i=0}^{b}\binom{b}{i}\frac{(-1)^i}{(a+i+1)^2} = \frac{a!\,b!}{(a+b+1)!}\sum_{k=1}^{b+1}\frac{1}{(a+k)}$ or,

$$\sum_{i=0}^{b}\binom{b}{i}\frac{(-1)^i}{(a+i+1)^2} = \frac{a!\,b!}{(a+b+1)!}(H_{a+b+1} - H_a).$$

Note that, by setting a=0 and b+1=m, identity derives to the well-known identity:

$$\sum_{p=1}^{m}(-1)^{p-1}\binom{m}{p}\frac{1}{p} = H_m.$$

And it finally holds that:

$$E[T_c] = nc\binom{n}{c}\sum_{i=0}^{c-1}\binom{c-1}{i}\frac{(-1)^i}{(n-c+1+i)^2} = nc\binom{n}{c}\frac{(n-c)!(c-1)!}{n!}\sum_{k=1}^{c}\frac{1}{(n-c+k)} = n(H_n - H_{n-c}).$$

# 10. Appendix: Inequalities on Subset Probabilities

We consider a non-EL distribution $\{p_i\}$, $0<p_i<1$, $1\leq i\leq N$, N is the support size.

First, from definition of $\{p_i\}$, it holds that: $\forall i : 1 > p_i > p_i^2 > p_i^3 \ldots$ and consequently:

$$1 = \sum_{i=1}^{N} p_i > \sum_{i=1}^{N} p_i^2 > \sum_{i=1}^{N} p_i^3 > \ldots$$

### (i) lemma 4 of [Anceaume14]

We give another proof of a very interesting lemma from [Anceaume14], lemma 4 pg7, which uses a proof of induction on N.

Let us divide the set $\{p_1,\ldots,p_N\}$ into subsets depending on the sign of $\left(p_i - \frac{1}{N}\right)$ and note $U = \{p_i \mid p_i > \frac{1}{N}\}$ and $V = \{p_i \mid p_i < \frac{1}{N}\}$. It holds that: $\sum_{i \in U}\left(p_i - \frac{1}{N}\right) + \sum_{j \in V}\left(p_j - \frac{1}{N}\right) = 0$, since all $p_i$ not belonging to U or V are equal to 1/N. Summation on U is positive and is the opposite of the summation on V. Also, note U is always non-empty otherwise V is also necessarily empty as well, which contradicts the fact that distribution is non-EL.

Then, noting that $\forall p_i \in U, \forall p_j \in V$ are such that $p_i > p_j$, it stands necessarily that:

$$\sum_{i \in U} \frac{1}{p_i}\left(p_i - \frac{1}{N}\right) + \sum_{j \in V} \frac{1}{p_j}\left(p_j - \frac{1}{N}\right) < 0, \text{ thus: } \sum_{i=1}^{N} \frac{1}{p_i}\left(p_i - \frac{1}{N}\right) < 0$$

which is [Anceaume14] result: $\sum_{i=1}^{N} \frac{1}{p_i} > N^2$. This result can be further generalized using the same argument with the following lemmas.

### (ii) Lemma 10.1

For any non-EL distribution $\{p_i\}$ and any non-decreasing function $f(x)$ on [0,1] such that $f(0) \neq f(1)$, then $\sum_l \left(p_l - \frac{1}{N}\right) f(p_l) > 0$. Inequality also holds when f verifies $f(a) > f(b)$ for any $0 < b < \frac{1}{N} < a < 1$. Reciprocally, if f is a non-increasing function $f(x)$ with $f(0) \neq f(1)$, or verifies $f(a) < f(b)$ for any $0 < b < \frac{1}{N} < a < 1$, then: $\sum_l \left(p_l - \frac{1}{N}\right) f(p_l) < 0$.

### (iii) Lemma 10.2

$$\sum_{i=1}^{N} p_i^k > \frac{1}{N^{k-1}}, \forall k \in \mathbb{Z} - \{0,1\}$$

Proof: Lemma 10.1 applies to $f(p_i) = \frac{1}{p_i^k}$, i.e., for any exponent k>0:

$$\sum_{i=1}^{N} \frac{1}{p_i^k}\left(p_i - \frac{1}{N}\right) < 0, \text{ so: } N \sum_{i=1}^{N} \frac{1}{p_i^{k-1}} < \sum_{i=1}^{N} \frac{1}{p_i^k}, \text{ hence: } \sum_{i=1}^{N} \frac{1}{p_i^k} > N^{k+1}.$$ On the other hand,

using a similar argument, it holds that: $\sum_{i=1}^{N} p_i^k > \frac{1}{N^{k-1}}$ for k>1. Hence same formula holds for any non-EL distribution and any exponent k, except for k=0 and 1 for which the two sides are equal. Note it is equivalent to another formulation: $\sum_{i=1}^{N} (Np_i)^k > N$.

A direct consequence is: $\sum_{i=1}^{N} \frac{1}{1-p_i} = \sum_{i=1}^{N} \sum_{k \geq 0} p_i^k = \sum_{k \geq 0} \sum_{i=1}^{N} p_i^k > N \sum_{k \geq 0} \frac{1}{N^k} = \frac{N^2}{N-1}$.

And similarly: $\sum_{i=1}^{N} \frac{p_i}{1-p_i} = \sum_{i=1}^{N} \sum_{k \geq 0} p_i^{k+1} > \frac{N}{N-1}$, this result is proved and used in Theorem 5 of [Anceaume14].

**(iv) Lemma 10.3** $\quad \sum_{i=1}^{N} p_i^k > \left( \sum_{i=1}^{N} p_i^2 \right)^{k-1}, \forall k > 2$

For any distribution it stands that: $\sum_{i=1}^{N} p_i \left( p_i - \sum_{l=1}^{N} p_l^2 \right) = 0$.

From that, by separating the set {p_1,...,p_N} into subsets depending on the sign of $\left( p_i - \sum_{i=1}^{N} p_i^2 \right)$, let us note $U = \{ p_i \mid p_i > \sum_{i=1}^{N} p_i^2 \}$ and $V = \{ p_i \mid p_i < \sum_{i=1}^{N} p_i^2 \}$. It holds that:

$\sum_{i \in U} p_i \left( p_i - \sum_{l=1}^{N} p_l^2 \right) + \sum_{j \in V} p_j \left( p_j - \sum_{l=1}^{N} p_l^2 \right) = 0$, i.e., summation on U is positive and is the opposite of the summation on V. Then noting that $\forall p_i \in U, \forall p_j \in V, p_i > p_j$, it stands necessarily that: $\sum_{i \in U} p_i^2 \left( p_i - \sum_{l=1}^{N} p_l^2 \right) + \sum_{j \in V} p_j^2 \left( p_j - \sum_{l=1}^{N} p_l^2 \right) > 0$.

Hence $\sum_{i=1}^{N} p_i^2 \left( p_i - \sum_{l=1}^{N} p_l^2 \right) > 0$ which readily implies $\sum_{i=1}^{N} p_i^3 > \left( \sum_{i=1}^{N} p_i^2 \right)^2$.

With the same argument it stands for k>2: $\sum_{i=1}^{N} p_i^{k-1} \left( p_i - \sum_{l=1}^{N} p_l^2 \right) > 0$. Hence $\sum_{i=1}^{N} p_i^k > \left( \sum_{l=1}^{N} p_l^{k-1} \right) \left( \sum_{l=1}^{N} p_l^2 \right)$ and using a simple induction argument it leads to $\sum_{i=1}^{N} p_i^k > \left( \sum_{i=1}^{N} p_i^2 \right)^{k-1}, \forall k > 2$. This completes the proof of the conjecture. QED.

Let us note also that from similar arguments, $\sum_{i \in U} \left( p_i - \sum_{l=1}^{N} p_l^2 \right) + \sum_{j \in V} \left( p_j - \sum_{l=1}^{N} p_l^2 \right) < 0$.

Hence $\sum_{i=1}^{N} \left( p_i - \sum_{l=1}^{N} p_l^2 \right) < 0$ which is another proof of: $\sum_{l=1}^{N} p_l^2 > \frac{1}{N}$.

A more general relation stems from two previous results: Both $\sum_{i=1}^{N} \frac{1}{1-p_i} > \frac{N^2}{N-1}$ and $\sum_{i=1}^{N} \frac{1}{p_i} > N^2$ are special cases, for j=1 and j=N-1, of next Lemma 10.4 with is a generalization to sums of subset probabilities.

**(v) Lemma 10.4**  $\sum_{|J|=j} \frac{1}{1-P_J} > \sum_{|J|=j} \frac{N}{N-j}, \ 0 < j < N$

Note that equality holds when j=0. Another formulation is: $\sum_{|J|=j} \frac{1}{P_J} > \sum_{|J|=j} \frac{N}{j}, \ 0 < j < N$, with equality for j=N. A proof goes like this:

We know ([Berthet17] Appendix3 relation 1) that $\sum_{|J|=j}(1-P_J) = \binom{N-1}{j}$ so we can consider the new distribution $\left\{(1-P_J)\binom{N-1}{j}^{-1}\right\}$ of $\binom{N}{j}$ possible elements (one per subset J). It is non-EL. And, since for any non-EL distribution $\{p_i\}$ with N elements: $\sum_{i=1}^{N} \frac{1}{p_i} > N^2$, then, for the new distribution: $\binom{N-1}{j} \sum_{|J|=j} \frac{1}{1-P_J} > \binom{N}{j}^2$ which implies for any non-EL distribution: $\sum_{|J|=j} \frac{1}{1-P_J} > \binom{N}{j} \frac{N}{N-j}, \ 0 < j < N$ .  QED

This expression can also be derived differently by noting that $\sum_{|J|=j} P_J = \binom{N-1}{j-1} = \sum_{|J|=j} \frac{j}{N}$, 1≤j≤N, or $\sum_{|J|=j}\left(P_J - \frac{j}{N}\right) = 0$. Using an argument similar to the one used in Lemma 10.1:

$\sum_{|J|=j; P_J > \frac{j}{N}} \left(P_J - \frac{j}{N}\right) + \sum_{|J|=j; P_J < \frac{j}{N}} \left(P_J - \frac{j}{N}\right) = 0$ necessarily implies that: $\sum_{|J|=j} P_J\left(P_J - \frac{j}{N}\right) > 0$

so $\sum_{|J|=j} P_J^2 > \sum_{|J|=j}\left(\frac{j}{N}\right)^2$. Similarly, $\sum_{|J|=j} P_J^k > \sum_{|J|=j}\left(\frac{j}{N}\right)^k$ for k>1. Equality stands for k=0 and k=1. Finally, summation over all k≥0, lends the desired relation.  QED

Let us mention the dual relation: $\sum_{|J|=j} \frac{1}{P_J^k} > \sum_{|J|=j} \left(\frac{N}{j}\right)^k$ for all 0<j≤N and k>0.

# 11. Appendix: Proof of $R_N^{N+2} = R_N^N \binom{N+2}{3} \frac{1}{4N}\left(3 + \sum_{i=1}^{N} p_i^2\right)$

Using WolframAlpha©, relation can be directly proved for N=2, 3, 4:

R(4,2): Simplify (x^4+(1-x)^4-1)/(2!*x*(1-x)*comb(4,3)*(3+x^2+(1-x)^2)/(4*2))  is -1.

R(5,3): Simplify (x^5+y^5+(1-x-y)^5-(x+y)^5-(1-x)^5-(1-y)^5+1)/(3!*x*y*(1-x-y)*comb(5,3)*(3+x^2+y^2+(1-x-y)^2)/(4*3))  is 1.

R(6,4): simplify (x^6+y^6+z^6+(1-x-y-z)^6-(x+y)^6-(x+z)^6-(y+z)^6-(1-x-y)^6-(1-x-z)^6-(1-y-z)^6+(1-x)^6+(1-y)^6+(1-z)^6+(x+y+z)^6-1)/(4!*x*y*z*(1-x-y-z)*comb(6,3)*(3+x^2+y^2+z^2+(1-x-y-z)^2)/(4*4))  is -1.

For the general case, we use the recurrence: $R_N^{N+2} = R_N^{N+1} - \sum_{l=1}^{N} p_l(1-p_l)^{N+1} R_{N-1,\{l\}}^{N+1}, \forall k > 0$

We assume the induction hypothesis holds on the distribution over (N-1) elements, i.e. the original distribution without element 'l', i.e.:

$$R_{N-1,\{l\}}^{N+1} = R_{N-1,\{l\}}^{N-1} \binom{N+1}{3} \frac{1}{4(N-1)}\left(\sum_{i=1, i \neq l}^{N}\left(\frac{p_i}{1-p_l}\right)^2 + 3\right).$$

Then: $R_N^{N+2} = R_N^{N+1} - \sum_{l=1}^{N} p_l(1-p_l)^{N+1} R_{N-1,\{l\}}^{N-1} \binom{N+1}{3} \frac{1}{4(N-1)}\left(\sum_{i=1, i \neq l}^{N}\left(\frac{p_i}{1-p_l}\right)^2 + 3\right)$

$\Leftrightarrow R_N^{N+2} = R_N^{N+1} + \sum_{l=1}^{N} (1-p_l)^2 R_N^N \binom{N+1}{3} \frac{1}{4N(N-1)}\left(\sum_{i=1, i \neq l}^{N}\left(\frac{p_i}{1-p_l}\right)^2 + 3\right)$

$\Leftrightarrow R_N^{N+2} = R_N^{N+1} + \sum_{l=1}^{N} \frac{N+1}{2} R_N^N \frac{1}{2 \cdot 3!}\left(\left(\sum_{i=1, i \neq l}^{N} p_i^2\right) + 3(1-p_l)^2\right)$

$\Leftrightarrow R_N^{N+2} = R_N^{N+1}\left(1 + \frac{1}{2 \cdot 3!}\sum_{l=1}^{N}\left(3(1-p_l)^2 - p_l^2 + \sum_{i=1}^{N} p_i^2\right)\right)$

Hence: $R_N^{N+2} = R_N^{N+1}\left(1 + \frac{1}{2 \cdot 3!}\left((N+2)\left(\sum_{i=1}^{N} p_i^2\right) + 3N - 6\right)\right)$

And finally: $R_N^{N+2} = R_N^{N+1} \frac{N+2}{2 \cdot 3!}\left(3 + \sum_{i=1}^{N} p_i^2\right) = R_N^N \binom{N+2}{3} \frac{1}{4N}\left(3 + \sum_{i=1}^{N} p_i^2\right)$  QED

# 12. Appendix: Minimality of EL for Waiting Time Expectation of a Partial Collection

A direct consequence of $\sum_{i=1}^{N}\frac{1}{1-p_i} = \sum_{k=0}^{+\infty}\sum_{i=1}^{N}p_i^k > \frac{N^2}{N-1}$ (see Appendix Section 10) is:

$E[T_2] > \frac{2N-1}{N-1} = N(H_N - H_{N-2})$. This means that EL and only EL is always minimal regarding the waiting time expectation for an incomplete subset of size 2. In order to extend this result to any partial collection size, we use the expression $I_J$, defined in [Berthet16] where J is a subset of size j on $\{1,..,N\}$:

$$I_J = \sum_{\substack{permutation \\ \{i_1,i_2,...i_j\} of J}} \frac{p_{i_1} p_{i_2} .. p_{i_j}}{(1-p_{i_1})(1-p_{i_1}-p_{i_2})..(1-p_{i_1}-p_{i_2}..-p_{i_j})}.$$

We are interested in the sum of $I_J$ over all subsets J of size j over a N-size support, noted $\sum_{|J|=j} I_J$. Expectation of a partial collection of size n verifies the relation $E[T_n] = \sum_{0\leq |J|<n} I_J$. This formulation of expectation was first expressed to our knowledge in [Ferrante12] with the help of conditional probabilities (see also [Ferrante14] and [Berthet16] for a proof of equivalence of this formula to the standard formula).

We want to prove that: $\sum_{|J|=j} I_J > \frac{N}{N-j}$, for all j, 1≤j<N.

It is fairly easy to check equality of the two terms for an EL distribution since in that case $I_J = \binom{N-1}{|J|}^{-1}$ and thus $\sum_{|J|=j} I_J = \binom{N}{j}\binom{N-1}{j}^{-1} = \frac{N}{N-j}$ [Berthet16]. Therefore strict inequality means that $\sum_{|J|=j} I_J$ is minimal if and only if the distribution is EL.

Proving the inequality implies that for any non-EL distribution: $E[T_n] > \sum_{0\leq j<n} \frac{N}{N-j}$ i.e.:

$E[T_n] > N(H_N - H_{N-n})$, and therefore that EL is minimal regarding the waiting time for a partial collection.

This proof is obvious for j=1 since: $\sum_{|J|=1} I_J = \sum_{i=1}^{N} \frac{p_i}{1-p_i}$ and, using lemma 10.2 in Appendix Section 10 : $\sum_{|J|=1} I_J > \frac{N}{N-1}$. In the general case we note that $\sum_{|J|=j} I_J$ is a sum over all permutations of all subsets of size j, then:

$$\sum_{|J|=j} I_J = \sum_{i_1=1}^{N}\sum_{\substack{i_2=1 \\ i_2\neq i_1}}^{N}...\sum_{\substack{i_j=1 \\ i_j\neq i_1\neq..\neq i_{j-1}}}^{N} \frac{p_{i_1} p_{i_2}...p_{i_j}}{(1-p_{i_1})(1-p_{i_1}-p_{i_2})..(1-p_{i_1}-p_{i_2}-..-p_{i_j})},$$ which actually is

[Ferrante12] notation. This expression can be rewritten:

$$\sum_{|J|=j} I_J = \sum_{i_1=1}^{N} \left( \frac{p_{i_1}}{1-p_{i_1}} \sum_{\substack{i_2=1 \\ i_2 \neq i_1}}^{N} \left( \frac{p_{i_2}}{1-p_{i_1}-p_{i_2}} .. \sum_{\substack{i_j=1 \\ i_j \neq i_1 \neq .. \neq i_{j-1}}}^{N} \frac{p_{i_j}}{1-p_{i_1}-p_{i_2}-..-p_{i_j}} \right) \right).$$

For j=2: $\sum_{|J|=2} I_J = \sum_{i=1}^{N} \sum_{\substack{j=1 \\ j \neq i}}^{N} \frac{p_i p_j}{(1-p_i)(1-p_i-p_j)} = \sum_{i=1}^{N} \left( \frac{p_i}{1-p_i} \sum_{\substack{j=1 \\ j \neq i}}^{N} \frac{p_j}{1-p_i-p_j} \right)$. We know that

for a non-EL N-size distribution $\{p_i\}$, $1 \leq i \leq N$: $\sum_{i=1}^{N} \frac{p_i}{1-p_i} > \frac{N}{N-1}$. Let us now consider the

set of distributions, one for each value of 'i', $1 \leq i \leq N$, on a (N-1)-size subset:

$\left\{ q_j = \frac{p_j}{1-p_i} \right\}$, $1 \leq j \leq N$, $j \neq i$. Clearly at least one of those is non-EL, and thus:

$\sum_{\substack{j=1 \\ j \neq i}}^{N} \frac{q_j}{1-q_j} = \sum_{\substack{j=1 \\ j \neq i}}^{N} \frac{p_j}{1-p_i-p_j} > \frac{N-1}{N-2}$. Finally: $\sum_{|J|=2} I_J > \sum_{i=1}^{N} \left( \frac{p_i}{1-p_i} \right) \frac{N-1}{N-2} > \frac{N}{N-2}$.

In a way similar to case j=2, we consider the distributions on a (N-(j-1)) size subset

$\left\{ q_{i_j} = \frac{p_{i_j}}{1-p_{i_1}-p_{i_2}-..-p_{i_{j-1}}} \right\}$ leading to: $\sum_{\substack{i_j=1 \\ i_j \neq i_1 \neq .. \neq i_{j-1}}}^{N} \frac{q_{i_j}}{1-q_{i_j}} > \frac{N-j+1}{N-j}$ and thus,

$\sum_{\substack{i_j=1 \\ i_j \neq i_1 \neq .. \neq i_{j-1}}}^{N} \frac{p_{i_j}}{1-p_{i_1}-p_{i_2}-..-p_{i_j}} > \frac{N-j+1}{N-j}$. Applied iteratively for each index from $i_j$ down to $i_1$, this yields the desired result: $\sum_{|J|=j} I_J > \frac{N}{N-1} \frac{N-1}{N-2} ... \frac{N-j+1}{N-j} = \frac{N}{N-j}$, and

consequently, minimality of EL regarding the waiting time of a partial collection.  QED

# 13. Appendix: Minimality of EL for Waiting Time CCDF

## (i) Comparison to EL regarding CCDF of a Complete Collection

We compare $R_N^k = \sum_{0 \leq |J| \leq N}(-1)^{|J|} P_J^k$, non-EL distributions and $EL_N^k = \sum_{0 \leq j \leq N}(-1)^j \binom{N}{j}\left(\frac{j}{N}\right)^k$ for the EL one, for all k≥N (both are null for 0≤k<N).

In the sequel we show that CCDF (and then expectation) of Waiting Time of complete collection for a non EL distribution is always higher than that of the EL case.

For a complete collection (size N) and a non-EL distribution, CCDF and expectation are:

$$P(C_N > k) = \sum_{j=0}^{N-1}(-1)^{N-1-j}\sum_{|J|=j} P_J^k = (-1)^{N-1}\left(R_N^k - (-1)^N\right) = 1 - (-1)^N R_N^k, \text{ and}$$

$$E[T_N] = \sum_{j=0}^{N-1}(-1)^{N-1-j}\sum_{|J|=j}\frac{1}{1-P_J} = \sum_{k \geq 0}(1 - (-1)^N R_N^k).$$

We want to prove that: $\forall k \geq N, (-1)^N R_N^k < (-1)^N EL_N^k$. For k<N, both sides are null.

First we consider the case k=N.

## (ii) EL and only EL is maximal for $(-1)^N R_N^N$

Inequality to prove for any non-EL distribution is: $(-1)^N R_N^N < \frac{N!}{N^N} = (-1)^N EL_N^N$.

Since $\sum_{|J|=j} P_J^2 = \binom{N-2}{j-2} + \binom{N-2}{j-1}\sum_{l=1}^{N} p_l^2$ (see Appendix 2 Relation 4 in [Berthet17]), then for

N=2, $R_2^2 = \sum_{0 \leq j \leq 2}(-1)^j \sum_{|J|=j} P_J^2 = -\sum_{|J|=1} P_J^2 + \sum_{|J|=2} P_J^2 = -\sum_{l=1}^{2} p_l^2 + 1$, hence $R_2^2 < \frac{1}{2}$ since $\sum_{l=1}^{2} p_l^2 > \frac{1}{2}$.

With N=3: $\sum_{|J|=j} P_J^3 = \binom{0}{j-3} + \binom{0}{j-2}\left(3\sum_{l=1}^{3} p_l^2 - \sum_{l=1}^{3} p_l^3\right) + \binom{0}{j-1}\sum_{l=1}^{3} p_l^3$, hence

$(-1)^3 R_3^3 = -\sum_{0 \leq j \leq 3}(-1)^j \sum_{|J|=j} P_J^3 = \sum_{l=1}^{3} p_l^3 - \left(3\sum_{l=1}^{3} p_l^2 - \sum_{l=1}^{3} p_l^3\right) + 1 = 1 - 3\sum_{l=1}^{3} p_l^2 + 2\sum_{l=1}^{3} p_l^3$,

using expression of $\sum_{|J|=j} P_J^3$ described in ([Berthet17], Relation 7).

It can be shown using WolframAlpha© that, for any (x,y): 0<x<1 and 0<y<1-x:

$1 - 3(x^2 + y^2 + (1-x-y)^2) + 2(x^3 + y^3 + (1-x-y)^3) < \frac{3!}{3^3} = \frac{2}{9}$.

Proof of the general case uses relation $R_N^N = (-1)^N N! \prod_{1 \le j \le N} p_j$ and next lemma.

### (iii) Lemma $\boxed{\prod_{1 \le j \le N}(Np_j) < 1}$

In other words, $\prod_{1 \le j \le N} p_j$ is maximal when $p_i = 1/N$. This can be proved by an induction argument on the size of the reference set as follows. First, $\prod_{1 \le j \le 2} p_j < \frac{1}{2^2}$ is true since $x(1-x) < \frac{1}{2^2}, \forall x, 0 < x < 1, x \ne 1/2$. Similarly, it stands that: $xy(1-x-y) < \frac{1}{3^3}$ for $\forall x, y: 0 < x < 1, 0 < y < 1-x, (x,y) \ne (\frac{1}{3}, \frac{1}{3})$. Induction at rank N gives (noting $x = p_{N+1}$): $\prod_{1 \le j \le N+1} p_j = \left(\prod_{1 \le j \le N} \frac{p_j}{1-x}\right)(1-x)^N x$, hence $\prod_{1 \le j \le N+1} p_j < \frac{1}{N^N}(1-x)^N x$. Finally, noting that $\forall x, 0 < x < 1, \forall N, N > 0: (1-x)^N x < \frac{N^N}{(N+1)^{N+1}}$ since positive function $(1-x)^N x$ has a single maximum equal to $\frac{N^N}{(N+1)^{N+1}}$ where its derivative is null (i.e. for $x = \frac{1}{N+1}$) completes the proof.

A direct consequence of $R_N^N$ and $EL_N^N$ definition is: $R_N^N = EL_N^N \cdot \prod_{1 \le j \le N}(Np_j)$ and then: $(-1)^N R_N^N < (-1)^N EL_N^N$.

Notice that, since it is known from a previous recurrence relation that $R_N^{N+1} = \frac{(N+1)}{2} R_N^N$, it immediately follows that: $(-1)^N R_N^{N+1} < (-1)^N EL_N^{N+1}$.

Next lemma is a generalization to n-size subsets of reference set and means that EL and only EL is maximal w.r.t. the LHS expression of the inequality.

### (iv) Lemma $\boxed{\forall n, 1 < n \le N, \sum_{|J|=n} \prod_{i \in J} p_i < \binom{N}{n} \frac{1}{N^n}}$

Equality stands for n=1, both sides being equal to 1. Case n=N corresponds to the previous lemma.

Inequality stands for n=2 since $\sum_{|J|=2} \prod_{i \in J} p_i = \frac{1}{2}\left(1 - \sum_{i=1}^{N} p_i^2\right) < \frac{N-1}{2N}$ thanks to Lemma 10.2 Appendix Section 10.

Proof in the general case is done by induction on the size of the reference set N, bearing in mind that lemma is obviously true for N=2.

Let us assume that lemma holds for any popularity on a reference set of size N and let us consider a popularity on a reference set of size N+1 and let us note $x = p_{N+1}$. Then:

$$\sum_{|J|=n} \prod_{i\in J} p_i = \sum_{|J|=n, x\in J} \prod_{i\in J} p_i + \sum_{|J|=n, x\notin J} \prod_{i\in J} p_i = x\left(\sum_{|J|=n-1, x\notin J} \prod_{i\in J} p_i\right) + \sum_{|J|=n, x\notin J} \prod_{i\in J} p_i$$

$$\Leftrightarrow \sum_{|J|=n} \prod_{i\in J} p_i = x(1-x)^{n-1}\left(\sum_{|J|=n-1, x\notin J} \prod_{i\in J} \frac{p_i}{1-x}\right) + (1-x)^n \sum_{|J|=n, x\notin J} \prod_{i\in J} \frac{p_i}{1-x}.$$

With induction hypothesis it holds that:

$$\sum_{|J|=n} \prod_{i\in J} p_i < x\binom{N}{n-1}\left(\frac{1-x}{N}\right)^{n-1} + \binom{N}{n}\left(\frac{1-x}{N}\right)^n = \left(\frac{1-x}{N}\right)^{n-1}\left(x\binom{N}{n-1} + \frac{1-x}{N}\binom{N}{n}\right)$$

Derivative of RHS positive function is null when

$$(n-1)\left(x\binom{N}{n-1} + \frac{1-x}{N}\binom{N}{n}\right) = (1-x)\left(\binom{N}{n-1} - \frac{1}{N}\binom{N}{n}\right), \text{ i.e., when } x = \frac{1}{N+1}.$$

Hence $\sum_{|J|=n} \prod_{i\in J} p_i < \left(\frac{1}{N+1}\right)^{n-1}\left(\frac{1}{N+1}\binom{N}{n-1} + \frac{1}{N+1}\binom{N}{n}\right) = \binom{N+1}{n}\left(\frac{1}{N+1}\right)^n$   QED

An amazing consequence is that for n=3, since $\sum_{|J|=3} \prod_{i\in J} p_i = \frac{1}{3}\left(\sum_{i=1}^{N} p_i^3 - \frac{3}{2}\sum_{i=1}^{N} p_i^2 + \frac{1}{2}\right)$, we obtain the relation: $2\sum_{i=1}^{N}\left(p_i^3 - \frac{1}{N^3}\right) < 3\sum_{i=1}^{N}\left(p_i^2 - \frac{1}{N^2}\right)$ which is stronger than the relation given at the end of Section 6.(iii).

To our knowledge, it is not known if similar relations exist for exponents higher than 3.

### (v) EL and only EL is maximal for $(-1)^N R_N^k$

For N=2 and k>2: $R_2^k = 1 - \sum_{l=1}^{2} p_l^k < 1 - \sum_{l=1}^{2}\left(\frac{1}{2}\right)^k$ since $\sum_{l=1}^{2} p_l^k > \frac{1}{2^{k-1}}$. Hence $R_2^k < EL_2^k$.

Note that for any distribution, $R_2^k$ is an increasing function of k, tending to 1, and similarly for $EL_2^k$.

For N=3, and k>3, $(-1)^3 R_3^k = 1 - \sum_{l=1}^{3}\left((1-p_l)^k - p_l^k\right)$ and $(-1)^3 EL_3^k = 1 - \frac{2^k - 1}{3^{k-1}}$. Proving that $\forall k, k > 3, \sum_{l=1}^{3}\left((1-p_l)^k - p_l^k\right) > \frac{2^k - 1}{3^{k-1}}$ is not trivial (WolframApha© does not converge).

General proof for k>N and any distribution, goes like this:

- Using a decomposition lemma introduced by [Anceaume14] page 2, (See also Relation 4 page 12 in [Berthet 17]): $\sum_{|J|=j} P_J^{k+1} = \sum_{l=1}^{n}\left(p_l \sum_{|J|=j,\, l \in J} P_J^k\right)$, $1 \le j \le n$, $0 \le k$, by a direct extension to complex sums, it stands that:

$R_N^k = \sum_{l=1}^{N} p_l \sum_{1 \le j \le N} (-1)^j \sum_{|J|=j-1,\, l \notin J} (p_l + P_J)^{k-1}$.

- Introduction of $\left\{\frac{p_j}{1-p_l}\right\}, j \in (1..N), j \ne l$ distribution with (N-1)

  elements: $R_N^k = \sum_{l=1}^{N} p_l (1-p_l)^{k-1} \sum_{1 \le j \le N} (-1)^j \sum_{|J|=j-1,\, l \notin J} \left(\frac{p_l}{1-p_l} + \frac{P_J}{1-p_l}\right)^{k-1}$

- Summation index change:

$R_N^k = \sum_{l=1}^{N} p_l (1-p_l)^{k-1} (-1) \sum_{0 \le j \le N-1} (-1)^j \sum_{|J|=j,\, l \notin J} \left(\frac{p_l}{1-p_l} + \frac{P_J}{1-p_l}\right)^{k-1}$

- We note $a = \frac{p_l}{1-p_l}$ which gives the binomial development (k≥N):

$\sum_{0 \le j \le N} (-1)^j \sum_{|J|=j} (a + P_J)^k = \sum_{u=N}^{k} a^{k-u} \binom{k}{u} \sum_{0 \le j \le N} (-1)^j \sum_{|J|=j} P_J^u$, since $\sum_{0 \le |J| \le N} (-1)^{|J|} P_J^u = 0$ for u<N.

- Hence, using notation introduced in section 5.(iv), we have:

$$R_N^k = \sum_{l=1}^{N} p_l (1-p_l)^{k-1} (-1) \sum_{u=N-1}^{k-1} \left(\frac{p_l}{1-p_l}\right)^{k-1-u} \binom{k-1}{u} R_{N-1,\{l\}}^u \;. \text{ Finally,}$$

$$(-1)^N R_N^k = (-1)^{N-1} \sum_{l=1}^{N} p_l^{\,k} \sum_{u=N-1}^{k-1} \left(\frac{1-p_l}{p_l}\right)^{u} \binom{k-1}{u} R_{N-1,\{l\}}^u$$

- Introducing strong induction hypothesis for any exponent u<k and any non-EL distribution $\left\{\frac{p_j}{1-p_l}\right\}, j \in (1..N), j \neq l$, hence: $(-1)^{N-1} R_{N-1,\{l\}}^u < (-1)^{N-1} EL_{N-1}^u$.

- Noting that $(-1)^N R_N^k$ is a sum of positive terms, each of them is bounded, then:

$$(-1)^N R_N^k < (-1)^{N-1} \sum_{l=1}^{N} p_l^{\,k} \sum_{u=N-1}^{k-1} \left(\frac{1-p_l}{p_l}\right)^{u} \binom{k-1}{u} \cdot EL_{N-1}^u$$

- Since equality stands for EL:

$$(-1)^N EL_N^k = (-1)^{N-1} \sum_{l=1}^{N} p_l^{\,k} \sum_{u=N-1}^{k-1} \left(\frac{1-p_l}{p_l}\right)^{u} \binom{k-1}{u} \cdot EL_{N-1}^u$$

Thus finally we have: $\boxed{(-1)^N R_N^k < (-1)^N EL_N^k, \forall k \geq N}$. QED.

Consequently, EL is maximal for the CDF form of $T_N$ probability (**complete** collection). A consequence is that EL is minimal for CCDF (and then expectation) of $T_N$ probability.

### (vi) Case of a Partial Collection

In the case of a partial collection, expressions of cumulative probability are more complex because binomial coefficients do not disappear as in the case of complete collections. However calculations can be carried out following the same scheme as for a complete collection.

The CCDF probability form of variable $T_n$ for a partial collection (n out of N) in k trials assuming a non-EL popularity is: $\Pr[T_n > k] = \sum_{j=0}^{n-1} (-1)^{n-1-j} \binom{N-j-1}{N-n} \sum_{|J|=j} P_J^{\,k}$. We want to prove that: $\forall k, k \geq n, \Pr[T_n > k] > \Pr[EL_n > k]$, where EL is a shorthand for the variable of the uniform case. We start with the simple cases n=2 and n=3.

For n= 2, we have: $\Pr[T_2 > k] = \sum_{j=0}^{1} (-1)^{1-j} \binom{N-j-1}{N-2} \sum_{|J|=j} P_J^{\,k} = \sum_{i=1}^{N} p_i^{\,k} > \frac{1}{N^{k-1}}$ thanks to Lemma 10.2 in Appendix Section 10, hence: $\Pr[T_2 > k] > \Pr[EL_2 > k]$.

For n=3, $\Pr[T_3 > k] = \sum_{j=0}^{2} (-1)^{2-j} \binom{N-j-1}{N-3} \sum_{|J|=j} P_J^{\,k} = -(N-2) \sum_{i=1}^{N} p_i^{\,k} + \sum_{1 \leq i < j \leq N} (p_i + p_j)^k$

Proving that EL is minimal for this expression is not trivial. WolframAlpha© does not converge even for N=4.

As mentioned previously, the sequence of operations detailed in the previous section can be used for the general case (partial collection with n<=N in k trials), and applied to $\Pr[T_n \leq k]$:

- Direct extension of [Anceaume14] decomposition Lemma to complex sums,

- Introduction of $\left\{\dfrac{p_j}{1-p_l}\right\}, j \in (1..N), j \neq l$ distribution with (N-1) elements,

- Summation index change,

- Binomial development (k≥N),

- strong induction hypothesis for any exponent u<k and any non-EL distribution $\left\{\dfrac{p_j}{1-p_l}\right\}, j \in (1..N), j \neq l$

- Noting that $\Pr[T_n \leq k]$ is a sum of positive terms, each of them is strictly bounded by the corresponding expression for EL.

Then keeping in mind that this derivation can be done with equalities for EL distribution, it follows that: $\Pr[T_n \leq k] < \Pr[EL_n \leq k]$, reminding that both are equal and null for k<n.

In conclusion, compared to any non-uniform distribution, EL is maximal for the CDF probability of a partial collection (and consequently EL is minimal for the CCDF and the expectation).

# 14. Appendix: Asymptotic Equivalence $WS^{-1}(j) \approx E[T_j]$

Following graph shows the function $\frac{1}{j}WS(E[T_j])$ for N=30 and different popularities, here power laws with skewness varying from a=1 (Zipf) to a=5 along with EL case and a lower bound which is given in the next sections. Computations are carried out using Flajolet expectation formula.

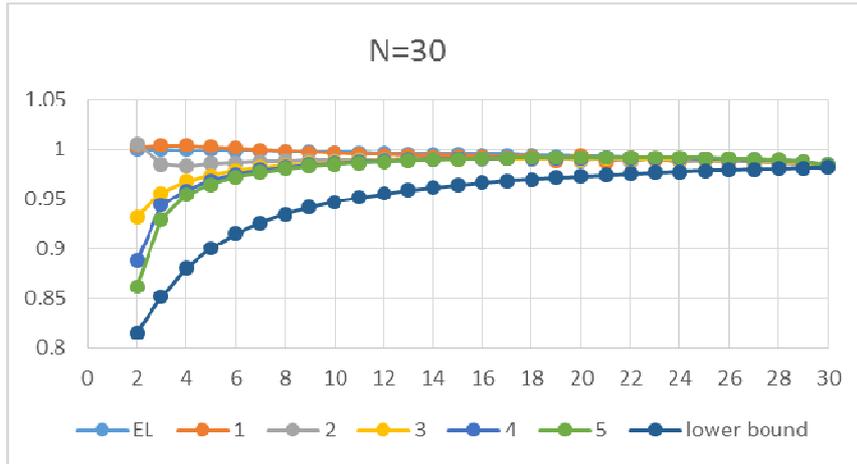

Obviously, for j=1, it stands whatever N, $WS(E[T_1]) = WS(1) = 1$, hence this case is ignored in the above graph.

Curves exhibit large differences when j=2 and few values afterwards. However, it is striking that, regardless of the skewness, curves tend to merge in a single one, actually that of EL. In particular, for j=N, curves converges towards EL value: $1-\left(1-\frac{1}{N}\right)^{NH_N}$.

When N>4, EL is such that: $N\left(1-\frac{1}{N}\right)^{NH_N} > e^{-1}$, hence: $\frac{1}{N}WS(E[T_N]) < 1 - \frac{1}{Ne}$

Using ln(1-x)~-x and $H_N$~ln(N)+γ, it is also $1-e^{-H_N} \approx 1-\frac{e^{-\gamma}}{N}$, (where $e^{-\gamma} = 0.56146$, this is the expressin given by [Boneh97] ) which tends to 1 when N→∞. It stands for EL in a similar way that: $\frac{WS(E[T_j])}{j} = \frac{N}{j}\left(1-\left(1-\frac{1}{N}\right)^{N(H_N-H_{N-j})}\right) \approx \frac{N}{j}\left(1-e^{-(H_N-H_{N-j})}\right)$.

Obviously, for a fixed j, $WS(E[T_j])$ can be seen as an increasing function of N.

Thus, when N→∞, limit of $\frac{WS(E[T_j])}{j}$ is: $\frac{N}{j}(1-\frac{N-j}{N}) = 1$.

**Analysis for N=j=2 and non-EL popularity**

In that case, popularity is parameterized with a single variable, say 'x'. It stands that:

$$E[T_2] = \sum_{k=1}^{2}(-1)^{k-1}\sum_{|J|=k}\frac{1}{P_J} = \sum_{i=1}^{2}\frac{1}{p_i} - 1, \text{ hence } E[T_2] = -1 + \frac{1}{x} + \frac{1}{1-x} \text{ where } 0<x<1.$$

Then, here and in the sequel, we make extensive use of the well-known relation:
$$\lim_{x\to 0^+}(1-x)^{\frac{1}{x}} = \frac{1}{e}, \text{ i.e. } \sim 0.36788.$$

Hence, expression $\sum_{i=1}^{2}(1-p_i)^{E[T_2]}$ has the following bounds: $\frac{1}{4} < x^{E[T_2]} + (1-x)^{E[T_2]} < \frac{1}{e}$.

The lower bound corresponds to EL case (i.e. for x=1/2) which is: $2\left(1-\frac{1}{2}\right)^{2H_2} = \frac{1}{4}$, and the upper bound to the case where an element of the distribution tends to 1 (and the other to 0).

Hence when N=2 and j=2, for any popularity (i.e. any value of x), it stands that:
$$1 - \frac{1}{8} > \frac{WS(E[T_2])}{2} > 1 - \frac{1}{2e}, \text{ where RHS term is } \sim 0.81606.$$

**Analysis for N=j=3**

For N=3, let us note the distribution {x,y,z} with x+y+z=1. Then:
$$E[T_3] = \sum_{k=1}^{3}(-1)^{k-1}\sum_{|J|=k}\frac{1}{P_J} = \frac{1}{x} + \frac{1}{y} + \frac{1}{z} - \frac{1}{x+y} - \frac{1}{x+z} - \frac{1}{y+z} + 1.$$

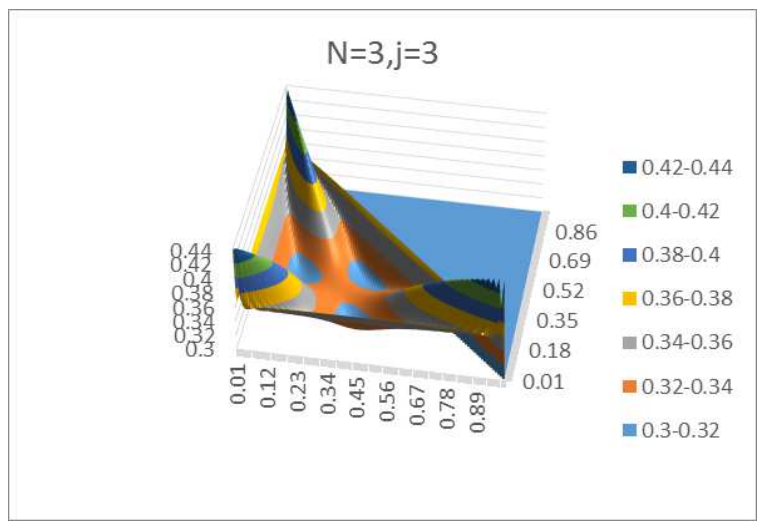

When one of (x,y,z) tends to 1, e.g. when x→1 (and identifying z to y and y→0),
$$E[T_3] \sim \frac{2}{y} - \frac{1}{2y} = \frac{3}{2y}, \text{ expression } (1-x)^{E[T_3]} + (1-y)^{E[T_3]} + (1-z)^{E[T_3]} \text{ is upper-bounded}$$

by: $\lim_{y\to 0} 2(1-y)^{\frac{3}{2y}} = 2e^{-\frac{3}{2}} \approx 0.44626$.

**Analysis for N=j=4**

With distribution {x,y,z,t} such that x+y+z+t=1,

$$E[T_4] = \sum_{k=1}^{4}(-1)^{k-1}\sum_{|J|=k}\frac{1}{P_J} = \frac{1}{x}+\frac{1}{y}+\frac{1}{z}+\frac{1}{t}-\frac{1}{x+y}-\frac{1}{x+z}-\frac{1}{x+t}-\frac{1}{y+z}-\frac{1}{y+t}-\frac{1}{z+t}$$
$$+\frac{1}{x+y+z}+\frac{1}{x+y+t}+\frac{1}{x+z+t}+\frac{1}{y+z+t}-1$$

Hence when x→1 (and identifying z and t to y, with y→0),

$$E[T_4] \sim \frac{1}{y}\left(3-\frac{3}{2}+\frac{1}{3}\right) = \frac{11}{6y} \text{ and } \lim_{y\to 0} 3\cdot(1-y)^{\frac{11}{6y}} = 3\cdot e^{-\frac{11}{6}} \approx 0.47964.$$

**Generalization to any complete collection (j=N)**

In the general case, from $E[T_N] = \sum_{k=1}^{N}(-1)^{k-1}\sum_{\substack{|J|=k \\ x\in J}}\frac{1}{P_J}+\sum_{k=1}^{N}(-1)^{k-1}\sum_{\substack{|J|=k \\ x\notin J}}\frac{1}{P_J}$, and letting x be an

element of the popularity tending to 1 (thus, others are identified to, say, y which tends to 0), we have: $E[T_N] \to \sum_{k=1}^{N}(-1)^{k-1}\sum_{\substack{|J|=k \\ x\in J}}1+\sum_{k=1}^{N}(-1)^{k-1}\sum_{\substack{|J|=k \\ x\notin J}}\frac{1}{k\cdot y} = \sum_{k=1}^{N}(-1)^{k-1}\binom{N-1}{k-1}+\frac{1}{y}\sum_{k=1}^{N}(-1)^{k-1}\frac{1}{k}\binom{N-1}{k}$

And, since first summation is null, finally: $E[T_N] \to \frac{H_{N-1}}{y}$.

Then $\lim_{y\to 0} N - WS(E[T_N]) = (N-1)\cdot\lim_{y\to 0}(1-y)^{\frac{H_{N-1}}{y}} = (N-1)\cdot e^{-H_{N-1}}$.

Observe that when N→∞, since $Ne^{-H_N} < e^{-\gamma}$, limit is upper bounded by $e^{-\gamma}$.

In conclusion, the following bound holds: $\frac{WS(E[T_N])}{N} > \frac{N-(N-1)\cdot e^{-H_{N-1}}}{N}$, and (WS is always upper-bounded by N) finally:

$$\boxed{1 > \frac{WS(E[T_N])}{N} > 1-\frac{e^{-\gamma}}{N}}.$$

**Note on Minimal value**

Although it is not necessary in our calculations, it is worth mentioning that surprisingly the minimum of $\sum_{i=1}^{N}(1-p_i)^{E[T_N]}$ is NOT the EL case when N>2. This is clearly visible on the (N=3, j=3) graph where the three minima of $\sum_{i=1}^{3}(1-p_i)^{E[T_3]}$ have a different color than the unique EL case. Indeed, EL expression is $3\left(1-\frac{1}{3}\right)^{3H_3} = 0.322567$, and we found three symmetrical minima at ~ 0.312403, one of them at (0.417, 0.417, 0.167).

Similarly for N=4, EL case is 0.363834 and a minimum at 0.326982712 is observed for (0.087  0.304  0.304  0.305).

It is not known whether an analytical solution exists for the minimas.

Worth also mentionning is that, for the distribution such that all but one element, say x, are equal, $\sum_{i=1}^{N}(1-p_i)^{E[T_N]}$ tends to $e^{-1}$ when x→0, and $e^{-\gamma}$ when x→1. When N increases to infinite, abciss of the minimum gets closer to 0, hence minimum tends to $e^{-1}$.

**Analysis for N=3, j=2**

We now consider the case of an incomplete collection. For N=3 and j=2, the following graph of $\sum_{i=1}^{3}(1-p_i)^{E[T_2]}$ is obtained:

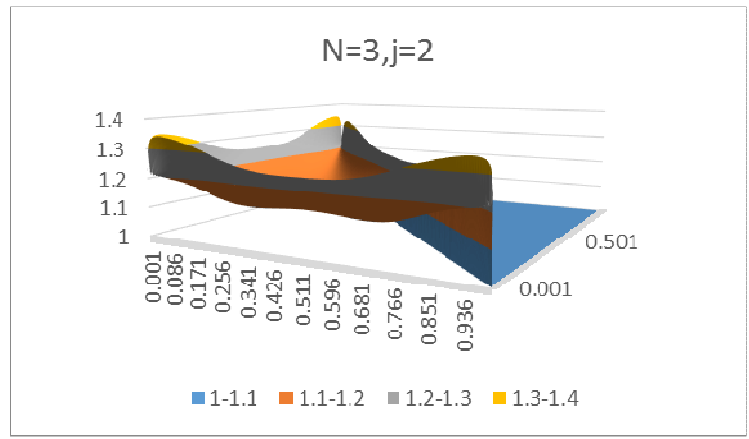

Let us note the distribution {x,y,z} with x+y+z=1. We have:

$E[T_2] = 1 + \dfrac{x}{1-x} + \dfrac{y}{1-y} + \dfrac{1-x-y}{x+y}$, then: $[E[T_2]]_{y=0} = 1 + \dfrac{x}{1-x} + \dfrac{1-x}{x}$.

Hence: $\left[(1-x)^{E[T_2]} + (1-y)^{E[T_2]} + (x+y)^{E[T_2]}\right]_{y=0} = (1-x)^{[E[T_2]]_{y=0}} + 1 + x^{[E[T_2]]_{y=0}}$

Finally: $\sum_{i=1}^{3}(1-p_i)^{E[T_2]} < 1 + \dfrac{1}{e}$, i.e. 1.3678, using same argument as before.

Let us note that here again, minimum is 1.0855, and it is not the EL case which is 1.0886.

**Analysis j=2, whatever N**

It stands: $E[T_2] = \sum_{k=N-1}^{N}(-1)^{k-N+1}\binom{k-1}{N-2}\sum_{|J|=k}\dfrac{1}{P_J} = \sum_{i=1}^{N}\dfrac{1}{1-p_i} - (N-1) = 1 + \sum_{i=1}^{N}\dfrac{p_i}{1-p_i}$, hence

$\sum_{i=1}^{N}(1-p_i)^{E[T_2]} = \sum_{i=1}^{N}(1-p_i)^{\left(1+\sum_{i=1}^{N}\frac{p_i}{1-p_i}\right)}$.

We extend the previous result obtained for N=3 by setting (N-2) items to 0, leading to $\sum_{i=1}^{N}(1-p_i)^{E[T_2]} < N - 2 + \dfrac{1}{e}$.

**Generalization**

An upper bound of $\sum_{i=1}^{N}(1-p_i)^{E[T_j]}$ is obtained when N-j elements are null, and the other elements are bounded by the value obtained for a complete collection.

The proof is as follows. Let us use Von Schelling notation:

$$E[T_j] = \sum_{k=N-j+1}^{N}(-1)^{j+k-N-1}\binom{k-1}{N-j}\sum_{|J|=k}\frac{1}{P_J},$$

and denote by x an element of the reference set,

then: $E[T_j] = \sum_{k=N-j+1}^{N}(-1)^{j+k-N-1}\binom{k-1}{N-j}\sum_{\substack{|J|=k\\x\in J}}\frac{1}{P_J} + \sum_{k=N-j+1}^{N-1}(-1)^{j+k-N-1}\binom{k-1}{N-j}\sum_{\substack{|J|=k\\x\notin J}}\frac{1}{P_J}.$

Let us assume x is null:

$$[E[T_j]]_{x=0} = \sum_{k=N-j+1}^{N}(-1)^{j+k-N-1}\binom{k-1}{N-j}\sum_{|J|=k-1}\frac{1}{P_J} + \sum_{k=N-j+1}^{N-1}(-1)^{j+k-N-1}\binom{k-1}{N-j}\sum_{|J|=k}\frac{1}{P_J},$$

With a variable change in the first summation (u=k-1) and extending the second summation to k=N-j: $[E[T_j]]_{x=0} = \sum_{u=N-j}^{N-1}(-1)^{j+u-N}\binom{u}{N-j}\sum_{|J|=u}\frac{1}{P_J} + \sum_{k=N-j}^{N-1}(-1)^{j+k-N-1}\binom{k-1}{N-j}\sum_{|J|=k}\frac{1}{P_J}.$

Then: $[E[T_j]]_{x=0} = \sum_{k=N-j}^{N-1}(-1)^{j+k-N}\binom{k-1}{N-j-1}\sum_{|J|=k}\frac{1}{P_J}$. RHS is the expression of waiting time expectation for a partial collection of size j among N-1 elements (after exclusion of element 'x'. The trick is that since 'x' is a null-weight element, the distribution remains the same for all the other elements).

Then: $\left[\sum_{i=1}^{N}(1-p_i)^{E[T_j]}\right]_{x=0} = 1 + \sum_{i=1}^{N-1}(1-p_i)^{[E[T_j]]_{x=0}}.$

Same reasoning can be applied iteratively for N-j variables, and, using the bound obtained for a complete collection (applied to j variables), we finally obtain:

$$\boxed{N-WS(E[T_j]) = \sum_{i=1}^{N}(1-p_i)^{E[T_j]} < N-j+(j-1)e^{-H_{j-1}}, \quad \forall\{p_i\}, i\in\{1,..N\}}.$$

**Analysis of particular case N=6**

We find the following values for N=6: (setting x, y,z,u,w to 0.00001 and varying them by steps of 0.01 with t=1-x-y-z-u-w) and comparison to the formula:

| j (N=6) | Max observed | $N - j + (j-1)e^{-H_{j-1}}$ | Min observed |
|---|---|---|---|
| 2 | 4.36348 | $4.36788 = 4 + e^{-1}$ | 3.98245 |
| 3 | 3.44042 | 3.44626 | 3.02642 |
| 4 | 2.47332 | 2.47964 | 2.07605 |
| 5 | 1.49181 | $1.49806 = 1 + 4e^{-H_4}$ | 1.15416 |
| 6 | 0.509713 | $0.509719 = 5e^{-H_5}$ | 0.33358 |

Coming back to the first graph of this section, we see that function $\frac{1}{j}WS(E[T_j])$ has the following lower bound: $\frac{1}{j}WS(E[T_j]) > \frac{1}{j}\left(j - (j-1)e^{-H_{j-1}}\right) = 1 - \frac{j-1}{j}e^{-H_{j-1}}$.

Let us remark that first this bound is independent of N and when j→∞ (and also N), $\frac{1}{j}WS(E[T_j]) > 1 - \frac{e^{-\gamma}}{j}$. Secondly, for j=N: $\frac{1}{N}WS(E[T_j]) > 1 - \frac{N-1}{N}e^{-H_{N-1}}$.

Finally, function is upper-bounded: $1 > \frac{1}{j}WS(E[T_j])$. This is a direct consequence of Boneh conjecture on the appearance of graphs, which implies that for any popularity there is necessarily a value above which, for any x: $E^{-1}[T_x] > WS(x)$. Then, let $j = E^{-1}[T_x]$, or $x = E[T_j]$, this implies: $j > WS(E[T_j])$.

In conclusion, for any popularity and for j large enough it stands that $\boxed{j > WS(E[T_j]) > j - e^{-\gamma}}$ and then: $\boxed{WS(E[T_j]) \approx j}$.                                  QED.

# 15. Appendix: Waiting Time Expectation of power-law popularities

Working Set Expectation has a computable expression when the popularity distribution is a power law with parameter 'a' (aka skewness): $p_i = \dfrac{1}{H_{N,a} \cdot i^a}$, where $H_{N,a}$ is the Nth generalized Harmonic number. We assume $p_i \ll 1$ holds for all i, i.e., $\ln(1-p_i) \sim -p_i$, and then $WS(D) = \sum_{i=1}^{N}(1-e^{-p_i D})$. Moving this expression to the continuous domain:

$$WS(D) = \int_0^N (1-e^{-\frac{1}{H_{N,a} x^a} D}) \cdot dx = \left[ x - \frac{(D/H_{N,a})^{\frac{1}{a}}}{a} \cdot \Gamma(-\frac{1}{a}, \frac{D/H_{N,a}}{x^a}) \right]_0^N = N - \frac{(D/H_{N,a})^{\frac{1}{a}}}{a} \cdot \Gamma(-\frac{1}{a}, \frac{D/H_{N,a}}{N^a})$$

where $\Gamma$ is the incomplete gamma function. This can be expressed with the 'generalized exponential integral' function: $E_p(z) = z^{p-1}\Gamma(1-p, z)$ (with a non-integer parameter p), hence: $\quad WS(D) = N(1 - \dfrac{1}{a} \cdot E_{1+\frac{1}{a}}(\dfrac{D}{H_{N,a} N^a}))$.

Consequently, its inverse function is: $\quad WS^{-1}(D) = H_{N,a} N^a E^{-1}_{1+\frac{1}{a}}(a(1-\dfrac{D}{N}))$.

We showed in Appendix Section 14 that, asymptotically, this expression is the Waiting Time expectation for a partial collection of size D.

This inverse function is not defined for D=N. But expectation of total collection can be approached as the limit of $WS^{-1}$ when N increases and D is N-c, for some constant c. This leads to the expression: $H_{N,a} N^a E^{-1}_{1+\frac{1}{a}}(\dfrac{ac}{N})$.

When N increases, reciprocal of Exponential Integral takes its value in the vicinity of 0+ and thus, can be approximated, regardless of p, by: $E_p^{-1}(x) = \ln(1/x) - \ln\ln(1/x)$, leading to: $H_{N,a} N^a \left( \ln(\dfrac{N}{ac}) - \ln\ln(\dfrac{N}{ac}) \right) \approx H_{N,a} N^a \ln N$.

This expression is equivalent to results given by Doumas and Papanicolaou [Doumas12].

Noting $\zeta(a) = \sum_{n=1}^{+\infty} \dfrac{1}{n^a} = \lim_{m \to +\infty} H_{m,a}$ the Riemann function which converges for a>1, asymptotic is $\zeta(a)N^a \ln(N)$ when a>1.

For a=1, since the Nth-harmonic number is $\ln(N)+\gamma+o(1/N)$, asymptotic is $H_N N \ln N \approx N \ln^2 N$. This result is also in [Flajolet92].

Finally, for 0<a<1, using approximation of generalized harmonic numbers $H_{N,a} \sim \dfrac{N^{1-a}}{1-a}$, asymptotic is $\dfrac{N \ln N}{1-a}$ [Doumas12].

(1) **Empirical verification of $WS^{-1}(n) \approx E[T_n]$ approximation for power laws**

Following graphs show how well $E[T_n]$ compare to $WS^{-1}(n)$ for power laws. Using GSL, we compare the inverse function of $WS(D) = N(1 - \frac{1}{a} \cdot E_{1+\frac{1}{a}}(\frac{D}{H_{N,a} N^a}))$, where a is the power-law parameter, with Flajolet&al. function: $E[T_n] = \sum_{q=0}^{n-1}(-1)^{n-1-q}\binom{N-q-1}{N-n}\sum_{|J|=q}\frac{1}{1-P_J}$

with $P_J = \sum_{i \in J} p_i$ and $p_i = \frac{1}{H_{N,a} \cdot i^a}$

We obtain the following graphs for power laws a =0.1 and a =1.0, assuming N=20:

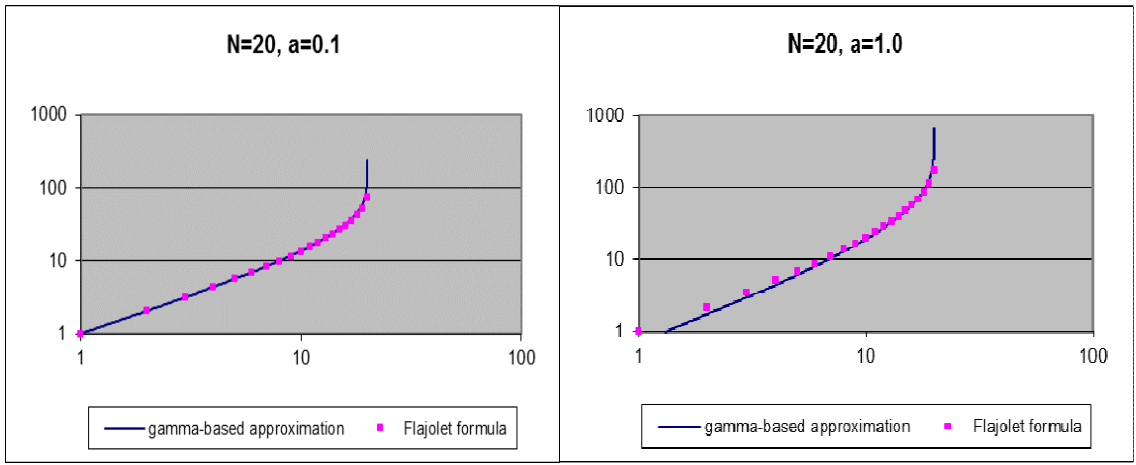

Fit is almost perfect for all points for a=0.1 (i.e. close to uniform). For a=1.0, comparison is valid for j above 10. This is likely due to the fact that the ln approximation (for all elements of the popularity, $\ln(1-p_i) = -p_i$), is less and less respected when the power-law parameter increases.

Of course, by definition, gamma-based $WS^{-1}(n)$ approximation tends to infinite for j=N whereas exact formula gives the defined value: $E[T_N] = \sum_{q=0}^{N-1}(-1)^{N-1-q}\sum_{|J|=q}\frac{1}{1-P_J}$.